\documentclass[12pt,journal,a4paper,draftcls,oneside,onecolumn]{IEEEtran}

\usepackage{times} 
\usepackage{graphicx}
\usepackage{subfigure}
\usepackage{cite}
\usepackage{citesort}
\usepackage{color}
\usepackage{psfrag}
\usepackage{amssymb}
\usepackage{epsfig}
\usepackage{pifont}
\usepackage{amsmath}
\usepackage{bm}
\usepackage{mathtools}
\usepackage{array}
\usepackage{multicol}
\usepackage[T1]{fontenc}
\usepackage{makecell}
\usepackage{algorithm}
\usepackage{algorithmic}
\usepackage{comment}

\newcommand{\fref}[1]{Fig.~\ref{#1}}
\newcommand{\tref}[1]{Table~\ref{#1}} 

\renewcommand{\baselinestretch}{1.20}

\makeatletter
\def\ifundefined{\@ifundefined}
\makeatother 

\begin{document}

\title{Augmented Affine Frequency Division Multiplexing for Both Low PAPR Signaling and Diversity Gain Protection}
\author{\IEEEauthorblockN{Zhou Lu, Mohammed El-Hajjar, \IEEEmembership{Senior Member,~IEEE,} and Lie-liang Yang, \IEEEmembership{Fellow,~IEEE, }}
\thanks{The authors are with the School of Electronics and Computer Science, University of Southampton, Southampton SO17 1BJ, United Kingdom (emails: zl15n22@soton.ac.uk, \{meh,lly\}@ecs.soton.ac.uk). The authors would like to acknowledge the financial support of the Engineering and Physical Sciences Research Council (EPSRC) projects under grants EP/X01228X/1, EP/X04047X/1 and EP/Y037243/1, as well as the Future Telecoms Research Hub, Platform for Driving Ultimate Connectivity (TITAN).}
}

%\vspace*{1cm}
\maketitle

\begin{abstract}
  Research results on Affine Frequency Division Multiplexing (AFDM) reveal that it experiences the same Peak-to-Average Power Ratio (PAPR) problem as conventional Orthogonal Frequency-Division Multiplexing (OFDM). On the other side, some references and also our studies demonstrate that AFDM involves an unneeded matrix, which is based on a parameter typically represented by $c_2$, for signalling. Hence, in this paper, an augmented AFDM scheme, referred to as A$^2$FDM, is proposed to mitigate the PAPR problem of AFDM, which is achieved by replacing the $c_2$ matrix in AFDM by a new unitary matrix that performs both sub-block-based Discrete Fourier Transform (DFT) and symbol mapping. Two symbol mapping schemes, namely interleaved mapping and localized mapping, are proposed for implementing A$^2$FDM, yielding the Interleaved A$^2$FDM and Localized A$^2$FDM. The input-output relationships of these schemes are derived and the complexity and the effects of system parameters on the performance of A$^2$FDM along with AFDM systems are analyzed. Furthermore, simulation results are provided to demonstrate and compare comprehensively the performance of the considered schemes in conjunction with different system settings and various operational conditions. Our studies and results demonstrate that, while A$^2$FDM is capable of circumventing the PAPR problem faced by AFDM, it is capable of attaining the achievable diversity gain, when AFDM is operated in its undesirable conditions resulting in the loss of the diversity gain available.
\end{abstract}
%%%%%%%%%%%%%%%%%%%%%%%%%%%%%%%%%%

\section{Introduction}\label{section-A2FDM-Introduction}

Next-generation wireless systems, including Beyond Fifth Generation (B5G) and Sixth Generation (6G) wireless networks, are expected to meet the demands of communications in high-mobility scenarios for high throughput, low latency, and ultra-massive connectivity~\cite{de2021survey}. Employing such capabilities is particularly crucial for the reliable communications on high-frequency bands in the environments, such as, Vehicle-to-Everything (V2X) networks, high-speed rail, and Unmanned Aerial Vehicle (UAV) systems~\cite{yang2020ultra}. As a typical classic multicarrier technique, Orthogonal Frequency Division Multiplexing (OFDM) has been widely adopted in the standardized wireless systems, which demonstrates robust performance in low to mild time-varying frequency-selective fading channels~\cite{bemani2023affine}. However, in high-mobility scenarios, the pronounced Doppler frequency shifts associated with multiple propagation paths exacerbate the time-varying multipath fading, leading to significant signal distortion, disrupting the orthogonality between subcarriers, and ultimately, degrading the achievable system performance reflected by throughput and reliability~\cite{li2022cross}. Moreover, OFDM struggles to achieve diversity gain in linear time-varying (LTV) channels, preventing it from achieving the optimal performance, especially, in high-mobility communications scenarios~\cite{hadani2017orthogonal}.

Orthogonal Time Frequency Space (OTFS), proposed by Hadani in 2017, has been regarded a prominent two-dimensional (2D) modulation scheme, which demonstrates strong resilience to Doppler frequency shifts while inheriting most of the advantages of OFDM~\cite{hadani2017orthogonal}. With the aid of the Zak transform, OTFS directly maps information symbols onto the 2D delay-Doppler grids, allowing superior performance in diverse communications conditions.  OTFS aided by effective equalization methods is capable of addressing the fading and time-varying challenges often encountered by OFDM, enabling full diversity provided in the time and frequency domains~\cite{wu2023message}. However, OTFS has some notable drawbacks, including the increased overhead as the result of its usage of the 2D transformations for pilot signalling and multi-user multiplexing~\cite{raviteja2019embedded}.

To address the limitations of OTFS and OFDM but take their respective advantages, Ali Bemani et al. introduced the Affine Frequency Division Multiplexing (AFDM) in 2021~\cite{bemani2021afdm,bemani2021affine}. Fundamentally, AFDM belongs to a multicarrier modulation approach innovatively built upon the framework of conventional OFDM. With AFDM, data symbols are transmitted using the orthogonal chirp signals generated from the Inverse Discrete Affine Fourier transform (IDAFT)~\cite{erseghe2005multicarrier}.  In contrast to OFDM and also Orthogonal Chirp Division Multiplexing (OCDM)~\cite{ouyang2016orthogonal}, AFDM offers enhanced flexibility via its tunable parameters. As shown in \cite{erseghe2005multicarrier,pei2000closed}, when appropriately set, these parameters allow for the efficient separation of the multipath components after the Discrete Affine Fourier transform (DAFT) at the receiver. Consequently, a comprehensive and sparse representation of the channel's delay-Doppler (DD) characteristics can be obtained, supplying potentially diversity gain. More specifically, via adapting the parameters of AFDM to the DD profile of wireless channel, AFDM is capable of spreading data symbols across the whole time-frequency plane. Hence, provided that a sufficiently powerful receiver equalizer is implemented to suppress the embedded interference, full diversity gain is available, which is comparable to the performance of OTFS~\cite{zhou2024overview}. Furthermore, AFDM enjoys almost-perfect
compatibility with the standard OFDM, and embraces OFDM and OCDM as the special signalling schemes within its framework~\cite{bemani2023affine}. The superb compatibility, plus its robustness and attractive performance in the highly time-varying channels, render AFDM as one of the promising candidate signaling waveforms for applications in high-mobility scenarios in the next generation wireless communications systems.

Research on AFDM is still in its early stages, which has already shown significant promise as a new waveform for high-mobility applications~\cite{bemani2023affine,savaux2023dft,tang2024time,yin2022pilot,benzine2023affine,ni2022afdm,bemani2024integrated,11185315,11082283}. More specifically, in literature, there are several studies that analyzed the adaptability of AFDM to various communication challenges. For instance, the modulation and demodulation techniques based on discrete Fourier transform (DFT) instead of DAFT have been studied in \cite{savaux2023dft}.  This allows AFDM to function as a precoded OFDM waveform, ensuing compatibility with the existing OFDM systems. To address the synchronization and channel estimation challenges in AFDM systems, the authors of \cite{tang2024time} proposed a maximum likelihood (ML) method, which leverages the redundant information contained in the chirp-periodic prefix (CPP) for more accurate synchronization. In \cite{yin2022pilot}, the authors introduced a pilot-assisted channel estimation method for AFDM systems, which allows to achieve the performance comparable to that by the systems with ideal channel state information (CSI). In the context of the time-varying sparse channels, the authors of \cite{benzine2023affine} compared the overhead required for channel estimation when various signalling waveforms are employed. It is shown that AFDM is capable of achieving the target mean square error (MSE) performance for channel estimation with a lower number of pilots and also a lower guard interval than the state-of-the-art single-carrier modulation (SCM), OFDM, and OTFS.  AFDM has also been investigated in terms of its potential for integrated sensing and communication (ISAC) applications~\cite{ni2022afdm,bemani2024integrated,ranasinghe2024joint}. Specifically, it is demonstrated in \cite{ni2022afdm} that AFDM is able to decouple the delay and Doppler effects within a relatively short time, and maintain robust sensing performance even when experiencing significant Doppler shifts. The comparative studies in \cite{bemani2024integrated} reveal that the ISAC systems built on OTFS, OCDM and AFDM, respectively, perform similarly for ranging and velocity estimation in terms of the MSE performance. Additionally, following OFDM, AFDM has been investigated with various index modulation schemes~\cite{10570960,10845819,10975107,11082283,11185315}, to reveal the design trade-off among spectral-efficiency, energy-efficiency and implementation complexity.

Despite its evident potentials, AFDM, like other multicarrier
communication systems, experiences the problem of high Peak-to-Average
Power Ratio
(PAPR)~\cite{rahmatallah2013peak,surabhi2019peak,wei2021charactering}. In
principle, AFDM signals have the same PAPR level as OFDM signals,
which is on the order of the number of subcarriers involved in
signalling~\cite{Lie-Liang-Resource-Allocation-book}. This elevated
PAPR can adversely affect communication efficiency and transceiver
design, including limiting communication coverage due to the
PAPR-resulted reduction of average transmit power, causing nonlinear
distortion - which results in out-of-band emission (OOBE) and
co-channel interference, and reducing the efficiency of both
digital-to-analog and analog-to-digital converters (DAC/ADC).  Since
AFDM shares a similar signalling structure as OFDM, many PAPR
mitigation techniques originally developed for OFDM can be adapted for
AFDM. However, most of these PAPR mitigation methods introduce
additional computational burden or/and processing steps, which yield
undesirable extra delay. Through reviewing the existing studies on
AFDM and our preliminary research, we recognize that the system
parameter $c_1$ involved in IDAFT/DAFT plays a crucial role for AFDM
systems to achieve diversity gain and, hence, improve the Bit Error
Rate (BER) performance. Conversely, the parameter $c_2$ in the
IDAFT/DAFT process has a negligible impact on the performance of AFDM
systems. Building on the above-mentioned observations, in this paper,
we propose an enhanced AFDM structure, namely, the Augmented AFDM
(A$^2$FDM), for solving the PAPR problem while maintaining the merits
of AFDM.  Our contributions are stated as follows:
\begin{itemize}

\item To mitigate the PAPR problem, we propose an A$^2$FDM signaling
  scheme by replacing the matrix in AFDM, which is based on the
  unneeded parameter $c_2$, with a new unitary matrix that performs
  both sub-block-based DFT and symbol mapping. Accordingly, two symbol
  mapping schemes are proposed, which we refer to as the interleaved
  A$^2$FDM (IA$^2$FDM) and localized A$^2$FDM (LA$^2$FDM).

\item We analyze in detail the input-output relationships of both
  IA$^2$FDM and LA$^2$FDM, showing the effect of the number,
  represented by a new system parameter $\mu$, of DFT sub-blocks on
  the PAPR and the achievable diversity order.

\item Considering specifically IA$^2$FDM, we analyze the effect of setting parameter $c_1$ on the performance of IA$^2$FDM along with AFDM, showing that the formula in references, such as \cite{bemani2023affine}, for setting this parameter is only necessary but not sufficient. Hence, a $c_1$ value following the formula in references may result in significant loss of diversity gain in AFDM systems. However, in IA$^2$FDM, similarly in LA$^2$FDM according to its principles, this problem can be avoided via the new parameter $\mu$, making the minimum diversity order be $\mu$, if full diversity gain is not attainable due to the use of an inappropriate parameter $c_1$.
  
 \item We analyze the PAPR of both IA$^2$FDM and LA$^2$FDM, revealing that both of them have the maximum PAPR on the order of $\mu$, instead of the total number of subcarriers in AFDM systems.

 \item We also analyze the complexity of A$^2$FDM and compare it with that of AFDM, showing that A$^2$FDM transmitter has only slightly added complexity from the operations of the sub-block-based DFT, while both IA$^2$FDM and AFDM receivers have the same complexity, when the same receiver equalizer is employed.

 \item Using Matlab simulations, a set of results are provided to comprehensively validate and compare the achievable performance of IA$^2$FDM, LA$^2$FDM and AFDM systems, as well as to demonstrate the impacts from the different parameters involved in these systems.
  
\end{itemize}

The remainder of this paper is organized as follows: Section~\ref{section-A2FDM-conventional AFDM structure} briefly introduces the principles of AFDM, and demonstrates the impacts of system parameters $c_1$ and $c_2$ on the performance of AFDM systems. Section~\ref{section-A2FDM-proposed A2FDM Structure} explains the principles of general A$^2$FDM, followed by that of the proposed IA$^2$FDM and LA$^2$FDM schemes, along with deriving the input-output relationships of the two special schemes. Section \ref{section-A2FDM-parameter analysis} presents the related analysis with respect to diversity, effect of parameters, PAPR, complexity, etc. Then, in Section \ref{section-A2FDM-simulation results}, performance results are demonstrated to compare the performance of various systems. Finally, in Section \ref{section-A2FDM-conclusion}, we conclude the research in this paper and present some future research directions.

\section{Affine Frequency-Division Multiplexing (AFDM) over Double-Selective Fading Channels} \label{section-A2FDM-conventional AFDM structure}

In this section, we review the principles of AFDM communicating over the time-frequency selective, i.e, double-selective, fading channels.

\subsection{Transmitter}\label{subsection-A2FDM-conventional AFDM structure-Modulation}

\begin{figure}[htbp]
    \centering
    \includegraphics[width = \linewidth]{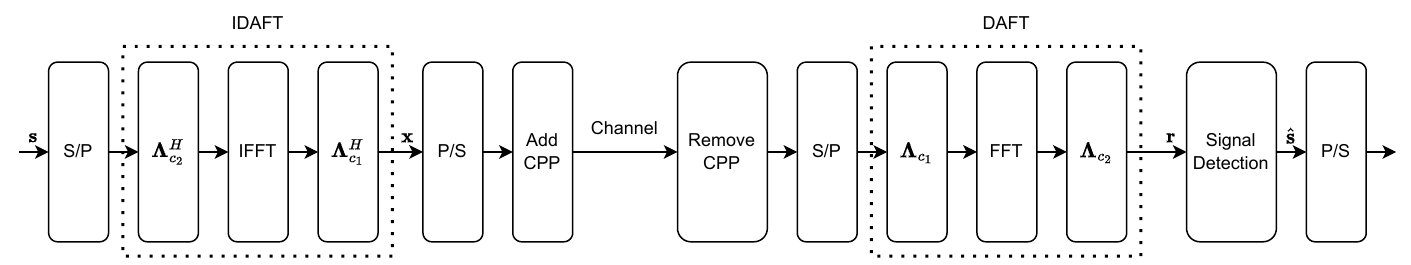}
    \caption{Block diagram for AFDM.}
    \label{fig:conventional AFDM}
\end{figure}

In AFDM~\cite{bemani2021afdm}, Discrete Affine Frequency Transform (DAFT) is the basic building block, which is the discrete version of AFT. In general, an AFDM transceiver has the structure as shown in \fref{fig:conventional AFDM}. Assume that the AFDM signalling uses $N$ subcarriers and let $\bm s \in \mathbb{C}^{N \times 1}$ denote the vector of information symbols generated by, such as, Quadrature Amplitude Modulation (QAM) or Phase-Shift Keying (PSK) modulation. Then, as shown in \fref{fig:conventional AFDM}, after the $N$-point inverse DAFT (IDAFT) on $\bm{s}$, the $N$ number of time-domain samples can be expressed as~\cite{bemani2023affine}
\begin{equation}\label{eq IDAFT}
    x_n = \frac{1}{\sqrt{N}}\sum^{N-1}_{m=0}e^{j2\pi \left(c_1n^2+\frac{mn}{N}+c_2m^2\right)}s_m,~n = 0,\cdots,N-1.
\end{equation}
Eq.~(\ref{eq IDAFT}) can be written in a product form as
\begin{equation}\label{eq:AFDM-h-2}
    x_n = e^{j2\pi c_1n^2}\times \left[\frac{1}{\sqrt{N}}\sum^{N-1}_{m=0} e^{j\frac{2\pi mn}{N}}\left(e^{j2\pi c_2m^2}\times s_m\right)\right],~~n = 0,\cdots,N-1.
\end{equation}
Comparing it with the time-domain samples in OFDM~\cite{yang2009multicarrier}, which is $x_n
=\frac{1}{\sqrt{N}}\displaystyle\sum^{N-1}_{m=0} e^{j\frac{2\pi
    mn}{N}}s_m$, $n = 0,\cdots,N-1$, we can understand that the two extra operations of AFDM on OFDM are as follows:
\begin{itemize}

  \item The $m$th subcarrier symbol $s_m$ is added an extra phase of $2\pi c_2m^2$, i.e., $s_m$ is preprocessed with a phase rotation of $2\pi c_2m^2$;
    \item The $n$th time-domain sample has an extra phase of $2\pi
      c_1n^2$. 
  
\end{itemize}
Alternatively, the exponential function in \eqref{eq IDAFT} can be expressed as
\begin{align}\label{eq:AFDM-h-2a}
e^{j2\pi \left(c_1n^2+\frac{mn}{N}+c_2m^2\right)}=e^{j \frac{2\pi(c_1nN+m+c_2m^2/n)n\Delta t}{N\Delta t}},
\end{align}
where $\Delta t$ is the sampling duration, which usually satisfies $\Delta t=1/B$ with $B$ being the total bandwidth of AFDM system. Then, we can readily realize that the subcarrier frequency of the $m$th subcarrier as
\begin{align}\label{eq:AFDM-h-2b}
  f_m(n)=&\frac{c_1nN+m+c_2m^2/n}{T_s},\nonumber\\
  =&\frac{m}{T_s}+\frac{c'_1n+c_2m^2/n}{T_s},n=0,1,\ldots,N-1;~m=0,1,\ldots,M-1,
\end{align}
where $T_s=N\Delta t$ is the symbol duration and $c'_1=c_1N$ is a constant, representing a modified system parameter of AFDM. Eq.~(\ref{eq:AFDM-h-2b}) explains that the frequency of the $m$th subcarrier is time-varying, increasing sample-by-sample. This time-varying subcarrier frequency enables a data symbol $s_m$, as seen in \eqref{eq IDAFT}, to be sent on a wide spectrum spanning many subcarriers - typically on $N$ subcarriers if the parameters $c_1$ and $c_2$ are appropriately selected, instead of always on the $m$th subcarrier with frequency $m/T_s$ in OFDM systems. During this transmission process, a data symbol $s_m$ can benefit from both the frequency-selectivity and time-selectivity of the wireless channel to achieve diversity gain, if a proper detection scheme is employed to mitigate the embedded interference.

When all time-domain samples are considered, Eq.~\eqref{eq IDAFT} can be written in matrix form as
\begin{equation}
    \bm x = \bm A^H\bm s = \bm \Lambda_{c_1}^H\bm F^H\bm \Lambda_{c_2}^H\bm s,
    \label{eq IDAFT matrix form}
\end{equation}
where $ \bm{x}=\left[x_0,x_1,\ldots,x_{N-1}\right]^T$, $\bm{A}=\bm{\Lambda}_{c_2}\bm{F} \bm{\Lambda}_{c_1}$ with $\bm F$ being the DFT matrix having entries $\frac{1}{\sqrt{N}}e^{j2\pi mn/N}$, and $\bm
\Lambda_c = \text{diag}(e^{-j2\pi cn^2}, n=0,1,\cdots,N-1)$ is a diagonal matrix, where $c = c_1$ or $c_2$.

Similar to OFDM, AFDM also needs a prefix to mitigate the effect of delay-spread due to multipath propagation. Owing to the property of chirp waveform, it has the periodicity of
\begin{equation}
    x_{n+uN} = e^{j2\pi c_1(u^2N^2+2uNn)}x_n,\quad u = 0,1,2,\cdots,
\end{equation}
where $u$ is the periodic index. Due to this special periodicity, the prefix, referred to as chirp-periodic prefix (CPP), in AFDM is given as
\begin{equation}
    x_{n} = e^{-j2\pi c_1(N^2+2Nn)}x_{n+N}, \quad n = -L_{CPP},\cdots,-1,
    \label{eq cpp}
\end{equation}
where $L_{CPP}$ is the length of CPP. Explicitly, if $c_1=0$, the CPP
is reduced to the CP in OFDM. Furthermore, similar to the requirement
by the CP in OFDM, the length of CPP $L_{CPP}$ should be an integer not
less than the channel delay spread. Moreover, as to be shown in the
forthcoming discourses, the introduction of CPP allows the
double-selective fading channel in AFDM to be converted to an
effectively periodic form that can help to obtain diversity gain.

\subsection{Double-Selective Fading Channel}\label{subsection-A2FDM-conventional AFDM structure-double selective fading channel}

After the parallel-to-serial (P/S) conversion, as seen in Fig.~\ref{fig:conventional AFDM}, the signal is transmitted over wireless channel, which may be double-selective and, hence, has the impulse response at time $n$ given by
\begin{equation}
    g_n(\ell) = \sum^L_{i=1}h_ie^{-j2\pi f_i n}\delta(\ell-\ell_i),
    \label{eq channel impulse response}
\end{equation}
where $L\geq 1$ is the number of propagation paths and $\delta(\cdot)$
is the Dirac delta function. In \eqref{eq channel impulse response},
$h_i$, $f_i$ and $\ell_i$ are complex gain, Doppler shift in digital
frequencies - which is the Doppler shift normalized by the sampling
frequency, and the integer delay, respectively, associated with the
$i$-th path. Specifically, if the sampling duration is $\Delta t=1/B$,
the sampling frequency is $B$. Hence, the actual Doppler shift is
approximately $f_iB=Nf_i\Delta f$, where $\Delta f$ is the subcarrier
spacing, satisfying $B=N\Delta f$. On the other hand, the actual delay of the $i$-th path is
about $\tau_i=\ell_i\Delta t$.

When $\bm{x}$ with its CPP is transmitted over the above described channel, the received time-domain samples can be represented as
\begin{equation}
    y_n = \sum^\infty_{\ell=0}x_{n-\ell}g_n(\ell)+w_n,\quad n = 0,1,\cdots,
    \label{eq received samples}
\end{equation}
where $w_n\sim \mathcal{CN}(0,N_0)$ is additive Gaussian noise.

Let us define $\nu_i \triangleq Nf_i = \alpha_i+\beta_i$. Hence,
$\nu_i \in [-\nu_{\max},\nu_{\max}]$ is the Doppler shift normalized
by the subcarrier spacing $\Delta f$, and $\nu_{\max}$ is the maximum
normalized Doppler shift. Accordingly, $\alpha_i\in
[-\alpha_{\max},\alpha_{\max}]$ is the integer part of $\nu_i$, and
$\beta_i$ is the fractional part of $\nu_i$ satisfying
$-\frac{1}{2}\leq\beta_i\leq\frac{1}{2}$. Assume that the maximum
delay of the channel satisfies $\ell_{\max}\triangleq \max(\ell_i)<N$,
and the CPP length $L_{CPP}$ is larger than $\ell_{\max}-1$. Then, the
matrix form of \eqref{eq received samples} can be written
as~\cite{bemani2023affine}
\begin{equation}
    \bm{y} = \bm{Hx}+\bm{w},
    \label{eq matrix form received signal}
\end{equation}
where $\bm{w} \sim \mathcal{CN}(\bm 0,N_0\bm I)$ is the additive white Gaussian noise vector, $\bm{I}$ is identity matrix, while the channel matrix $\bm H$ is given by~\cite{bemani2023affine}
\begin{equation}
    \bm H = \sum^L_{i=1}h_i\bm \Gamma_{CPP_i} \bm \Delta_{f_i} \bm \Pi^{\ell_i}.
    \label{eq channel matrix}
\end{equation}
Following \cite{bemani2023affine}, the terms in \eqref{eq channel matrix} are defined as follows. $\bm \Gamma_{CPP_i}$ is an $N\times
N$ diagonal matrix given as
\begin{equation}
    \bm \Gamma_{CPP_i} = \text{diag}\left(
    \begin{cases}
        e^{-j2\pi c_1(N^2-2N(\ell_i-n))},& n<\ell_i,\\
        1,&n\geq \ell_i,
    \end{cases},\\
    \quad n = 0,1,\cdots,N-1 \right),
\end{equation}
which, for $n<\ell_i$, has the contribution of the phase shifts induced by the CPP. When $n\geq \ell_i$, CPP has no effect on the signal, and hence the diagonal elements of $\bm \Gamma_{CPP_i}$ are equal to $1$.  $\bm \Delta_{f_i} \triangleq \text{diag}(e^{-j2\pi
  f_in},n = 0,1,\cdots,N-1)$ accounts for the effect of the Doppler shifts, while $\bm\Pi^{\ell_i}$ explains the effect of delay $\ell_i$ of the $\ell_i$th path, where $\bm \Pi$ is a forward cyclic-shift matrix expressed as~\cite{bemani2021afdm}
\begin{equation}
    \bm \Pi = 
    \begin{bmatrix}
        0&\cdots&0&1\\
        1&\cdots&0&0\\
        \vdots&\ddots&\ddots&\vdots\\
        0&\cdots&1&0
    \end{bmatrix}_{N\times N}.
\end{equation}

\subsection{Demodulation}\label{subsection-A2FDM-conventional AFDM structure-Demodulation}

As shown in \fref{fig:conventional AFDM}, for the receiver to demodulate the information symbols, the DAFT is firstly carried out on the received signal in \eqref{eq matrix form received signal}, which can be expressed as
\begin{equation}
\begin{aligned}
    \bm r &= \bm{Ay} =  \bm A\left(\sum^L_{i=1}h_i\bm \Gamma_{CPP_i} \bm \Delta_{f_i} \bm \Pi^{\ell_i}\right)\bm A^H\bm s+\bm w' \\
    &= \sum_{i=1}^Lh_i\bm{H}_{eff}^{(i)}\bm s+\bm w'\\
    &= \bm H_{eff}\bm s+\bm w',
\end{aligned}
  \label{eq demodulation}  
\end{equation}
where, by definition, $\bm H_{eff}^{(i)} = \bm{A}\bm \Gamma_{CPP_i}
\bm \Delta_{f_i} \bm \Pi^{\ell_i}\bm{A}^H$, $\bm H_{eff} =
\bm{A}\bm{H}\bm{A}^H$ accounts for the effective end-to-end channel,
where $\pmb{H}=\sum^L_{i=1}h_i\bm \Gamma_{CPP_i} \bm \Delta_{f_i} \bm
\Pi^{\ell_i}$, and $\bm w' = \bm{Aw}$.

In \eqref{eq demodulation}, $\bm{H}_{eff}$ is not a diagonal matrix, and the columns of $\bm{H}_{eff}$ are also not orthogonal. Hence, the symbols in $\bm{s}$ interfere with each other. Consequently, more complex equalization than that in the conventional OFDM systems is required.  In literature, various detection schemes have been proposed and studied~\cite{bemani2022low,9880774,10711268}, including MRC-based detection and pilot aided detection. Specifically, when a linear minimum mean-square error (MMSE) equalizer is employed, the decision variables can be formed as
\begin{align}%\label{eq:A2FDM-15}
\bm{y}=\bm{W}^H\bm{r},
 \end{align}
where the weight matrix $\bm{W}$ can be derived from \eqref{eq demodulation} and is expressed as
\begin{equation}   
\label{eq:A2FDM-15}
\bm{W}=\left(\bm{H}_{eff}\bm{H}_{eff}^H+\frac{1}{\gamma}\bm{I}\right)^{-1}\bm{H}_{eff},
\end{equation}
where $\gamma$ is the signal-to-noise ratio (SNR) per symbol.
\begin{table}[htbp]
    \centering
    \caption{Parameters for simulations of AFDM and A$^2$FDM.}
    \begin{tabular}{|c|c|} \hline 
         Parameter& Value\\ \hline 
 Light speed&$c = 3\times 10^8$~m/s\\\hline 
         Carrier frequency& $f_c = 3.5$~GHz\\ \hline 
         Subcarrier spacing& $\Delta f = 30$~kHz\\ \hline 
 Velocity of user&$v = 100$~kilometers/hour (km/h)\\ \hline  
         Number of subcarriers& $N=256$\\ \hline 
         Number of channel paths& $L = 10$\\ \hline 
         Modulation& $4$QAM\\ \hline
    \end{tabular}
    \label{tab:system parameters}
\end{table}

\begin{figure}[htbp]
    \centering
    \includegraphics[width = 0.65\linewidth]{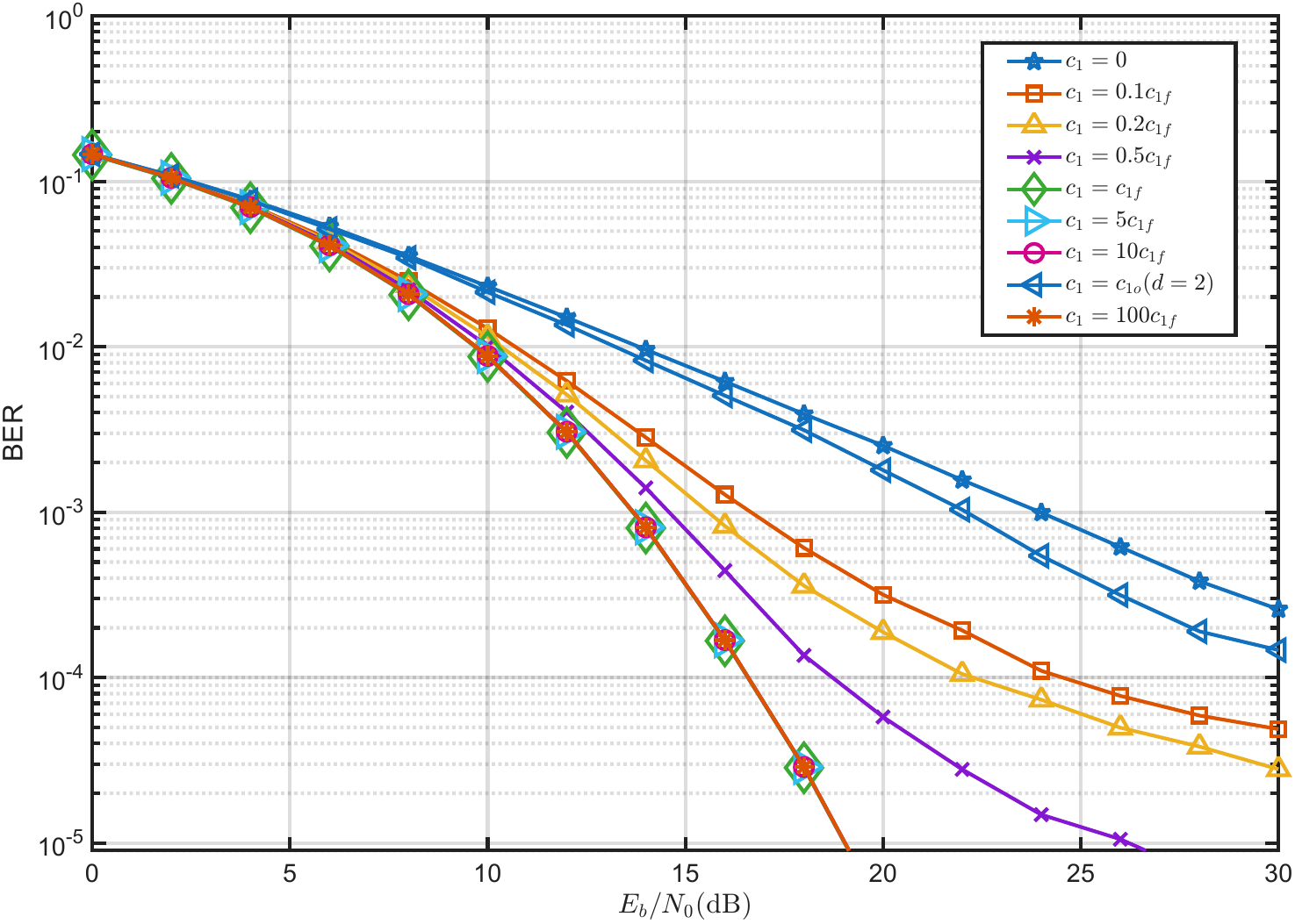}
    \caption{BER performance of AFDM system with MMSE equalization, with the settings of  $f_c = 3.5$GHz, $\Delta f = 30$~kHz, $v = 100$~km/h, $N=256$, $L=10$, $4$QAM, and $c_2 = 0$.}
    \label{fig:AFDM-with-different-pb}
\end{figure}

The researches~\cite{bemani2021afdm,bemani2023affine,bemani2021affine} show that the system parameter $c_1$ imposes a big impact on the achievable performance of AFDM systems. Specifically, to obtain the full diversity gain, $c_1$ should be chosen a value of
 \begin{equation}\label{eq:optimum-c1}
 c_1 > \left(c_{1f}= \frac{2\nu_{\max}}{2N \times \min(|\ell_j - \ell_i|)} \right),
 \end{equation}
where $c_{1f}$ is the minimum value of $c_1$ required for the AFDM system to achieve full diversity.  This is also illustrated in \fref{fig:AFDM-with-different-pb}, which compares the BER performance of the same AFDM system but with different values for the parameter $c_1$. The simulation parameters are detailed in Table~\ref{tab:system parameters}, unless otherwise specified.  In our simulations, all channel paths are assumed to experience Rayleigh fading with a common power $1/L$.

 \begin{figure}[htbp]
    \centering
    \includegraphics[width=0.65\linewidth]{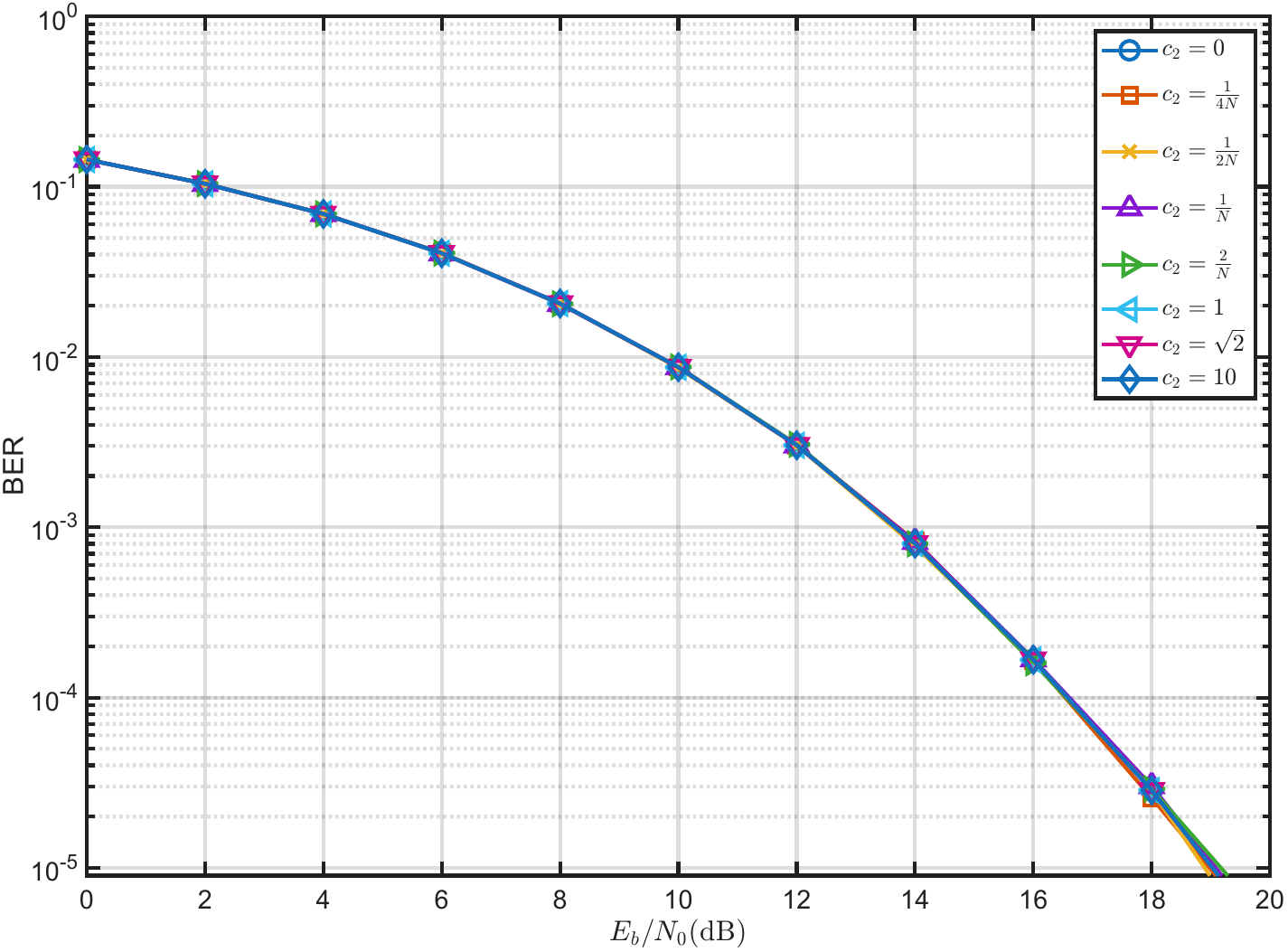}
    \caption{BER performance of AFDM systems with MMSE equalization and $c_1 = c_{1f}$.}
    \label{fig:ber c2}
\end{figure}

\begin{figure}[htbp]
    \centering
    \includegraphics[width=0.65\linewidth]{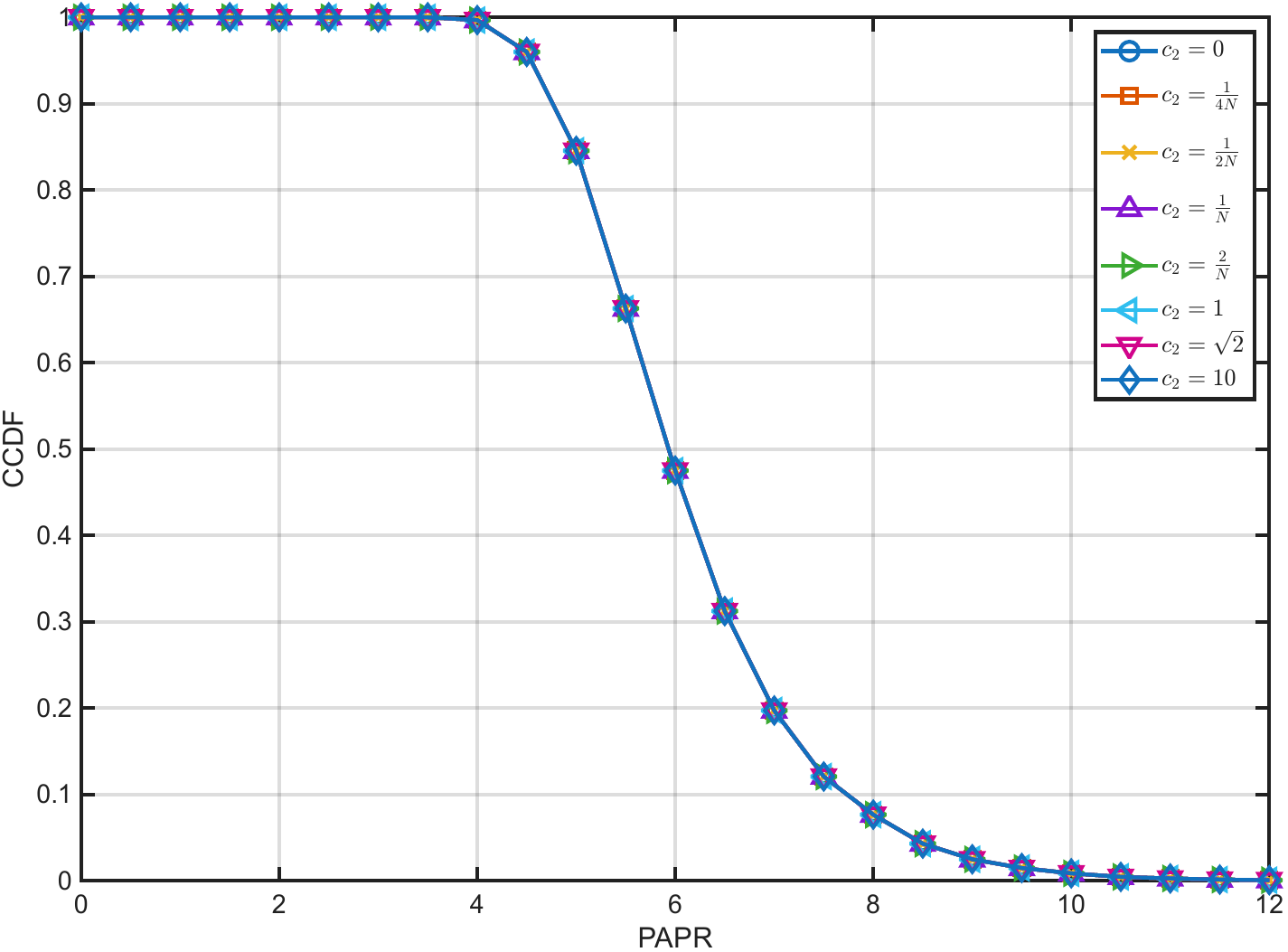}
    \caption{PAPR of AFDM transmit signals with $c_1 = c_{1f}$.}
    \label{fig:papr c2}
\end{figure}
The results in \fref{fig:AFDM-with-different-pb} show that the BER
performance of AFDM is strongly relied on the value of $c_1$.  The BER
performance improves with the increase of $c_1$ until
$c_1=c_{1f}$. Then, further increasing $c_1$ has little effect on the
BER performance of the AFDM system, unless a big but inappropriate
$c_1$ value, as shown by the curve corresponding to
``$c_1=c_{1o}(d=2)$'', which will be addressed in
Section~\ref{subsection-A2FDM-parameter analysis-c1 for diversity
  optimization}.

In contrast to $c_1$ having a big impact on the BER performance of
AFDM systems, the parameter $c_2$ yields little effect, which has
already been realized in some references. For example, in
\cite{bemani2023affine}, it is noted that optimum diversity order can
be achieved, provided that $c_2$ is chosen to be an irrational number
or a small rational number.  Moreover, the parameter $c_2$ is simply
ignored in the transmitter design in
\cite{erseghe2005multicarrier}. To illustrate the insignificance of
parameter $c_2$, we present \fref{fig:ber c2} and \fref{fig:papr c2}
to show its effect on the BER and PAPR performance, respectively, of
AFDM systems. The main parameters are the same as that shown in
Table~\ref{tab:system parameters}. As shown in \fref{fig:ber c2}, for
a given value of $c_1= c_{1f}$, different values of $c_2$ generate nearly the
same performance. It is worth noting that the BER performance is
nearly the same, even when $c_2$ is not small but as big as $10$. The
PAPR results in \fref{fig:papr c2} show that $c_2$ has no effect at
all, no matter what value it is set. Therefore, in general, the
diagonal matrix contributed by parameter $c_2$ can be removed from
AFDM systems. Note that, as the application of $c_1$ only shifts
signals in the frequency-domain, as to be shown by the analysis in
Section~\ref{subsubsection-A2FDM-proposed A2FDM Structure-design of
  DA2FT-IDA2FT}, it has no effect on PAPR.

However, AFDM experiences the same PAPR problem as OFDM, which should
be mitigated, especially, when uplink communications are
implemented. Accordingly, in this paper, we enhance the design of AFDM
to not only attain full diversity gain, but also significantly reduce
the PAPR of transmit signals.

\section{Augmented AFDM (A$^2$FDM)}\label{section-A2FDM-proposed A2FDM Structure}

We propose the A$^2$FDM to serve two purposes. The first is to reduce
the PAPR conflicted by AFDM, as above-mentioned. In AFDM, especially,
when the number of channel paths is relatively small, the selection of
parameter $c_1$ is dependent on the knowledge about the delays and
Doppler shifts, as seen in Eqs. (45) - (48) in
\cite{bemani2023affine}, which may not be always available in
practice. Furthermore, in practical mobile communications, delays and
Doppler shifts as well as their distributions are time-varying,
meaning that a fixed value of $c_1$ may never be optimal. Therefore,
it is important to set $c_1$ a value that is robust in different
communication environments. Hence, the second purpose is to enable
A$^2$FDM systems to achieve the near-optimum performance in terms of
diversity gain (orders) for any randomly selected $c_1$ value.

The fundamental principle of A$^2$FDM is to replace the $c_2$-related
diagonal matrix in AFDM by a block-diagonal matrix with component DFT
matrices, as detailed in the following subsection.

\subsection{Transmitter of A$^2$FDM}\label{subsection-A2FDM-proposed A2FDM Structure-design of DA2FT}
\begin{figure}[htbp]
    \centering
    \includegraphics[width=\linewidth]{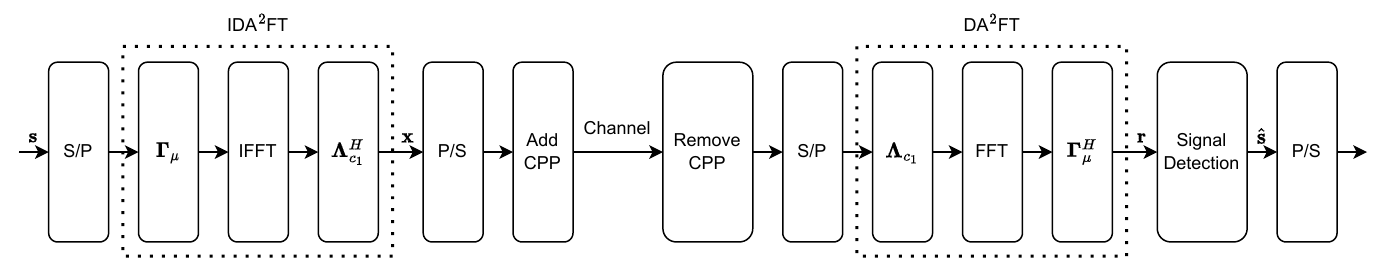}
    \caption{Block diagram of A$^2$FDM.}
    \label{fig:A2FDM}
\end{figure}

Along with OFDM, although various PAPR reduction methods have been
proposed~\cite{ochiai2002performance,bauml1996reducing,cimini2000peak,li1997effects,rahmatallah2013peak},
they usually introduce significant extra processes, which are not
desirable. Hence, to reduce the PAPR of AFDM and further enhance its
flexibility in implementation, but without significant increase of
implementation cost, we introduce an augmented AFDM, referred to as
A$^2$FDM, scheme, by exploiting the principles of single-carrier
frequency-division multiple-access
(SC-FDMA)~\cite{yang2009multicarrier}.

Specifically, as shown in \fref{fig:A2FDM}, instead of the DAFT in
AFDM, our A$^2$FDM scheme relies on a so-called discrete augmented
affine Fourier transform (DA$^2$FT) with the operator expressed as
\begin{equation}\label{Block DAFT}
    \bm A = \bm{\Gamma}^H_\mu\bm{F}\bm{\Lambda}_{c_1},
\end{equation}
where $\bm{\Gamma}_\mu$ is chosen to make $\bm{A}$ a unitary matrix,
i.e., satisfying $\bm{A}\bm{A}^H=\bm{A}^H\bm{A}=\bm{I}$. Since in this
paper, one of our design objectives is to solve the PAPR problem of
AFDM, following the principles of SC-FDMA \cite{yang2009multicarrier},
we design $\bm{\Gamma}_\mu$ as
\begin{align}\label{eq:A2FDM-18}
  \bm{\Gamma}_\mu =\bm{P\Upsilon}_\mu.
  \end{align}
 Again, assume that the total number of subcarriers is $N$. Then, in
 \eqref{eq:A2FDM-18}, $\bm{\Upsilon}_\mu$ is a $N\times N$
 block-diagonal matrix consisting of $\mu$ number of component FFT
 matrices of $N_\mu \times N_\mu$ dimensions, i.e.,

\begin{equation}\label{eq:A2FDM-19}
    \begin{aligned}
               \bm{\Upsilon}_\mu  &= \text{diag}\big(\underbrace{ \bm{F}_{N_\mu}, \bm{F}_{N_\mu}, \cdots, \bm{F}_{N_\mu}}_\mu\big)\\
            &=\bm I_\mu \otimes \bm F_{N_\mu},
    \end{aligned}
\end{equation}
where $\otimes$ is Kronecker product, $\bm{F}_{N_\mu}$ is a
$N_\mu\times N_\mu$ normalized FFT matrix, $N_\mu = N/\mu$ is an
integer, preferably equalling to $2^i$ for convenience of FFT
implementation. In \eqref{eq:A2FDM-18}, $\bm{P}$ is a $N\times N$
permutation matrix implementing the mapping of signals to different
subcarriers, so that signal power is redistributed to attain a lower
PAPR after IFFT. Following the interleaved FDMA and localized FDMA in
SC-FDMA~\cite{yang2009multicarrier}, in this paper, both the
interleaved permutation and localized permutation are considered,
resulting in the Interleaved A$^2$FDM (IA$^2$FDM) and Localized
A$^2$FDM (LA$^2$FDM) schemes. Since $\bm{P}$ is a permutation matrix,
it satisfies $\bm{P}^T\bm{P}=\bm{P}\bm{P}^T=\bm{I}$. Therefore, we can
readily show that DA$^2$FT matrix defined in \eqref{Block DAFT} is a
unitary matrix.

Note that in \eqref{eq:A2FDM-19}, we assumed that all the component
FFT matrices in $\bm{\Upsilon}_\mu$ have the same dimensions, for the
sake of simplicity of the forthcoming analysis. However, it is worth
noting that its extension to including the component FFT matrices of
various sizes is straightforward. This is practically the case, when a
base-station (BS) sends information to its downlink users at different
rates, or when different uplink users communicate with a BS at
different rates. In both cases, different users may be assigned
different amount of resources, such as different numbers of
subcarriers.

Corresponding to the structure of $\bm \Upsilon_\mu$, the $N$ data
symbols, such as QAM/PSK symbols, to be transmitted are divided into
$\mu$ groups of each having $N_\mu$ symbols, expressed as $\bm{s} =
[\bm{s}^T_0,\bm{s}^T_1,\cdots,\bm{s}^T_{\mu-1}]^T$ .  Let us express
$\bm{z} = \bm{\Upsilon}_\mu\bm{s}$. Then, according to $\bm{s}$,
$\bm{z}$ can be divided into $\bm{z} =
[\bm{z}^T_0,\bm{z}^T_1,\cdots,\bm{z}^T_{\mu-1}]^T$, where $\bm{z}_k$
is a $N_\mu$-length vector given by
\begin{equation}
    \bm{z}_k = \bm{F}_{N_\mu}\bm{s}_k, \quad k = 0,1,\cdots,\mu-1.
\end{equation}
More specifically, the $p$-th element of $\bm{z}_{k}$ is given by
\begin{equation}
    z_{kp} = \frac{1}{\sqrt{N_\mu}}\sum^{N_\mu-1}_{l=0}s_{kl}e^{-j2\pi\frac{pl}{N_\mu}}, \quad p = 0,1,\cdots,N_\mu-1,
    \label{eq elements form of z_k,p}
\end{equation}
where symbol $s_{kl}$ is the $l$-th element in $\bm{s}_k$, which is the symbol of $\bm{s}_{k N_\mu+l}$ in $\bm{s}$ for $k=0,1,\ldots,\mu-1$ and $l=0,1,\ldots,N_\mu-1$. Hence, the entries in $\bm{z} = \bm{\Upsilon}_\mu\bm{s}$ can be expressed as
\begin{equation}
\begin{aligned}
    z_{kN_\mu+p} &= \frac{1}{\sqrt{N_\mu}}\sum^{N_\mu-1}_{l=0}s_{k N_\mu+l}e^{-j2\pi\frac{(kN_\mu+p)(kN_\mu+l)}{N_\mu}} \\
    &=\frac{1}{\sqrt{N_\mu}}\sum^{N_\mu-1}_{l=0}s_{k N_\mu+l}e^{-j2\pi\frac{pl}{N_\mu}}
\end{aligned}
    \label{eq elements form of z}
\end{equation} 
for all $p = 0,1,\cdots,N_\mu-1$ and $k = 0,1,\cdots,\mu-1$.

Following the FFT, symbol permutation is carried out via multiplying matrix $\bm P$ on $\bm{z}$. Hence, different permutation strategies may be implemented via designing $\bm{P}$.  Following the IFDMA and LFDMA in SC-FDMA \cite{yang2009multicarrier}, in this paper, we propose the concepts of the IA$^2$FDM and LA$^2$FDM, which are detailed in the following sections.

\subsubsection{IA$^2$FDM}\label{subsubsection-A2FDM-proposed A2FDM Structure-design of DA2FT-IDA2FT}

With IA$^2$FDM, the $N_\mu$ symbols in $\bm{z}_k$ are evenly mapped via permutation onto $N_\mu$ subcarriers of the whole $N=\mu N_\mu$ subcarriers. After the subcarrier mapping, the transmitter forms an extended $N$-length symbol vector. Specifically, corresponding to $\bm{z}_k$, the mapping operation obeys
\begin{align}\label{eq:A2FDM-23}
  \bm{g}=& [g_{0},g_{1},\cdots,g_{N-1}]^T\nonumber\\
  =&\bm{P}_k\bm{z}_k,
\end{align}
where $\bm{P}_k$ is responsible for the mapping of the entries in
$\bm{z}_k$, which constitutes the $N_\mu$ columns of $\bm{P}$ that
match to the positions of $\bm{z}_k$ in $\bm{z}$.  Let us set
$\bm{z}_k=\left[z_{k0},z_{k1},\ldots,z_{k(N_\mu-1)}\right]^T$ for
$k=0,1,\ldots,\mu-1$. Then, when IA$^2$FDM is implemented, $\bm{P}_k$
is designed to set the value of ${g}_{m}$ according to
\begin{equation}
g_{m} = 
    \begin{cases}
        z_{kp},& m=p\mu+k,\\
        0,& \text{otherwise}.
    \end{cases}
\end{equation}
Hence, to implement the interleaved permutation, the columns of the
permutation matrix $\bm{P}$ can be constructed as
\begin{equation}\label{eq:A2FDM-25}
\bm{P}(:,kN_\mu+p)=\bm{I}_N(:,p\mu+k), 
\end{equation}
for $p = 0,1,\cdots,N_\mu-1$ and $k = 0,1,\cdots,\mu-1$, where $\bm{I}_N$ is a $N \times N$ identity matrix. Let $\bm g= \bm{Pz}=\bm
\Gamma_\mu \bm s$. Then, according to \eqref{eq:A2FDM-25}, the element at $kN_\mu+p$ in $\bm{z}$ is mapped to the position $p\mu+k$ in $\bm
g$, i.e., $g_{p\mu+k} = z_{kN_\mu+p}$ for $p = 0,1,\cdots,N_\mu-1$ and $k = 0,1,\cdots,\mu-1$.

\begin{figure}[htbp]
    \centering
    \includegraphics[width=0.5\linewidth]{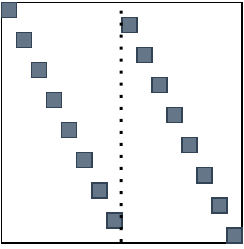}
    \caption{Structure of matrix $\bm P$ for IA$^2$FDM with $\mu=2$. }
    \label{fig interleaved mapping}
\end{figure}
An example of $\bm P$ implementing interleaved permutation with $N=16$
and $\mu=2$ is shown in \fref{fig interleaved mapping}, where a shaded
box represents an element $1$ and, otherwise, the element value is
$0$. As shown in \fref{fig interleaved mapping}, the symbols from $2$
groups are alternatively allocated to the $16$ subcarriers.

After the permutation operation, the remaining transmitter operations in IA$^2$FDM are the same as that in AFDM, including the IFFT, the phase shifting operation invoking parameter $c_1$, adding CPP, etc. It can be shown that the $n$th sample in the transmitted signal $\bm x$ can be expressed as
\begin{equation}
\begin{aligned}
    x_n &= \frac{1}{\sqrt{N}}\sum^{N-1}_{p\mu+k=0}g_{p\mu+k}e^{j2\pi\left(c_1n^2+\frac{(p\mu+k)n}{N}\right)}\\
    &=\frac{1}{\sqrt{N}}e^{j2\pi c_1n^2} \sum^{N-1}_{p\mu+k=0} z_{kN_\mu+p} e^{j2\pi \frac{(p\mu+k)n}{N}}\\
    &=\frac{1}{\sqrt{N}}e^{j2\pi c_1n^2} \sum^{\mu-1}_{k=0}e^{j2\pi\frac{kn}{N}} \sum^{N_\mu-1}_{p=0} z_{kN_\mu+p} e^{j2\pi\frac{pn}{N_\mu}}, \quad n = 0,1,\cdots,N-1.\\
\end{aligned}
\label{eq transmitted signal xn IDA2FT}
\end{equation}
Denote $n = aN_\mu+b$ with $a = 0,1,\cdots,\mu-1$ and $b =
0,1,\cdots,N_\mu-1$. Then, \eqref{eq transmitted signal xn IDA2FT} can be written as
\begin{equation}
\begin{aligned}
    x_{(aN_\mu+b)}&=\frac{1}{\sqrt{N}}e^{j2\pi c_1(aN_\mu+b)^2} \sum^{\mu-1}_{k=0}e^{j2\pi\frac{k(aN_\mu+b)}{N}} \sum^{N_\mu-1}_{p=0} z_{kN_\mu+p} e^{j2\pi\frac{p(aN_\mu+b)}{N_\mu}}\\
    &=\frac{1}{\sqrt{\mu}}e^{j2\pi c_1(aN_\mu+b)^2} \sum^{\mu-1}_{k=0}e^{j2\pi\frac{k(aN_\mu+b)}{N}}\frac{1}{\sqrt{N_\mu}}\sum^{N_\mu-1}_{p=0} z_{kN_\mu+p} e^{j2\pi\frac{paN_\mu+pb}{N_\mu}}\\
    &=\frac{1}{\sqrt{\mu}}e^{j2\pi c_1(aN_\mu+b)^2} \sum^{\mu-1}_{k=0}e^{j2\pi\frac{k(aN_\mu+b)}{N}}s_{kN_\mu+b}.
\end{aligned}
\label{eq transmitted signal x_{a N_mu+b} IDA2FT}
\end{equation}
Hence, when substituting $n = aN_\mu+b$ back into \eqref{eq transmitted signal x_{a N_mu+b} IDA2FT} and denoting $m=kN_\mu+b$, the time-domain samples can be expressed as
\begin{equation}
    x_n = \frac{1}{\sqrt{\mu}}e^{j2\pi c_1n^2} \sum^{\mu-1}_{k=0}s_me^{j2\pi\frac{k n}{N}},~n = 0,1,\cdots,N-1.
    \label{eq xn with IDA2FT}
\end{equation}
Accordingly, the continuous-time signal can be expressed as
\begin{equation}\label{eq:A2FDM-29}
    x(t) = \sqrt{\frac{P}{\mu}}\sum^{\mu-1}_{k=0}s_me^{j2\pi\frac{(c_1Nn+k)t }{T_s}},
\end{equation}
where $P$ is transmit power and $T_s$ is symbol duration. \eqref{eq:A2FDM-29} shows that, at any time, there are only $\mu$ active subcarriers, which includes the frequencies given by $(c_1Nn+k)/T_s$, determined by both $k$ and $c_1Nn$. Therefore, the PAPR of A$^2$FDM can be significantly lower than that of AFDM.

\subsubsection{LA$^2$FDM}\label{subsubsection-A2FDM-proposed A2FDM Structure-design of DA2FT-LDA2FT}
When the LA$^2$FDM is implemented, the $N_\mu$ symbols in $\bm{z}_k$ obtained from the $N_\mu$-point FFT processing are assigned to $N_\mu$ successive subcarriers, which can be formulated as
\begin{equation}
{g}_{m} = 
    \begin{cases}
        z_{kp},& m=kN_\mu+p,~p=0,1,\ldots,N_\mu-1\\
        0,& \text{otherwise},
    \end{cases}
\end{equation}
Hence, matrix $\bm{P}$ is simply an identity matrix $\bm{I}_N$, and $\bm{P}_k$ is just constituted by the $N_\mu$ columns corresponding to $\bm{z}_k$, yielding
\begin{equation}
         \quad g_{kN_\mu+p} = z_{kN_\mu+p},~p = 0,1,\cdots,N_\mu-1;~k = 0,1,\cdots,\mu-1.
    \label{eq G expression}
\end{equation}
Therefore, in LA$^2$FDM, the samples in $\bm{x}$ can be expressed as
\begin{equation}
\begin{aligned}
    x_n &= \frac{1}{\sqrt{N}}\sum^{N-1}_{kN_\mu+p=0}g_{kN_\mu+p}e^{j2\pi\left(c_1n^2+\frac{(kN_\mu+p)n}{N}\right)}\\
    &=\frac{1}{\sqrt{N}}e^{j2\pi c_1n^2} \sum^{N-1}_{kN_\mu+p=0} z_{kN_\mu+p} e^{j2\pi \frac{(kN_\mu+p)n}{N}}\\
    &=\frac{1}{\sqrt{N}}e^{j2\pi c_1n^2} \sum^{\mu-1}_{k=0}e^{j2\pi\frac{kn}{\mu}} \sum^{N_\mu-1}_{p=0} z_{kN_\mu+p}  e^{j2\pi\frac{pn}{N}}.
\end{aligned}
\label{eq transmitted signal xn LDA2FT}
\end{equation}
Let us define $n = b\mu+a$ for $b = 0,1,\cdots,N_\mu-1$ and $a =
0,1,\cdots,\mu-1$. Then, we have
\begin{equation}
\begin{aligned}
    x_{(b\mu+a)} &=\frac{1}{\sqrt{N}}e^{j2\pi c_1(b\mu+a)^2} \sum^{\mu-1}_{k=0} e^{j2\pi\frac{k (b\mu+a)}{\mu}}\sum^{N_\mu-1}_{p=0} z_{kN_\mu+p}  e^{j2\pi\frac{p(b\mu+a)}{N}}\\
    &=\frac{1}{\sqrt{\mu}}e^{j2\pi c_1(b\mu+a)^2} \sum^{\mu-1}_{k=0}e^{j2\pi\frac{k a}{\mu}} \frac{1}{\sqrt{N_\mu}}\sum^{N_\mu-1}_{p=0} z_{kN_\mu+p}  e^{j2\pi\frac{p(b\mu +a)}{N}}\\
    &=\frac{1}{\sqrt{\mu}}e^{j2\pi c_1(b\mu+a)^2} \sum^{\mu-1}_{k=0}e^{j2\pi\frac{k a}{\mu}} x'_{(b\mu+a)},
\end{aligned}
\label{eq transmitted signal x_{b mu+a} LDA2FT}
\end{equation}
where we defined $x'_{(b\mu+a)}=\frac{1}{\sqrt{N_\mu}}\sum^{N_\mu-1}_{p=0} z_{kN_\mu+p}
e^{j2\pi\frac{p(b\mu +a)}{N}}$, which is analyzed in Appendix~\ref{Appendix-A2FDM-1}, showing that 
\begin{equation}
x'_n=
    \begin{cases}
        s_{kN_\mu+b},&n=b\mu~(a=0),\\
        \frac{1}{N_\mu}\sum^{N_\mu-1}_{l=0}s_{k N_\mu+l}\left(\frac{1-e^{j\frac{2\pi n}{\mu}}}{1-e^{j\frac{2\pi n}{N}}e^{-j\frac{2\pi l}{N_\mu}}}\right), &\text{otherwise}~(a\neq 0),
    \end{cases}
    \label{eq x'n summary}
\end{equation}
where $b = 0,1,\cdots,N_\mu-1$.

Upon substituting \eqref{eq x'n summary} into \eqref{eq transmitted signal x_{b mu+a} LDA2FT}, the time-domain samples transmitted in LA$^2$FDM are given by
\begin{equation}
x_n=
\begin{cases}
    \frac{1}{\sqrt{\mu}}e^{j2\pi c_1n^2} \sum^{\mu-1}_{k=0}e^{j2\pi\frac{k [n/\mu]}{\mu}}s_m,&n=b\mu,\\
    \frac{1}{N_\mu\sqrt{\mu}}e^{j2\pi c_1n^2} \sum^{\mu-1}_{k=0}e^{j2\pi\frac{k [n/\mu]}{\mu}}\sum^{N_\mu-1}_{l=0}s_m\left(\frac{1-e^{j\frac{2\pi n}{\mu}}}{1-e^{j\frac{2\pi n}{N}}e^{-j\frac{2\pi l}{N_\mu}}}\right), &n\neq b\mu,
\end{cases}
\label{eq:A2FDM-35}
\end{equation}
where $[n/\mu]$ denotes the $\mu$-modulo operation on $n$, and $m =
kN_\mu+b=0,1,\cdots,N-1$, when $b$ and $k$ scan the ranges of $b =
0,1,\cdots,N_\mu$ and $k = 0,1,\cdots,\mu-1$.

Following the IA$^2$FDM in Section~\ref{subsubsection-A2FDM-proposed A2FDM Structure-design of DA2FT-IDA2FT}, it can be analyzed that at the times $t=b\mu\Delta t$, there are only $\mu$ subcarriers activated at frequencies $(c_1nN+kN_\mu)/T_s$ for $k=0,1,\ldots,\mu$. At the other times of $t\neq b\mu\Delta t$, there are more than $\mu$ active subcarriers. Specifically, for a given $k$, the group of $N_\mu$ symbols are transmitted on three subcarriers located at $(c_1nN+kN_\mu)/T_s$, $N_\mu/T_s$ and $\mu/T_s$. Therefore, the total number of subcarriers activated is $\mu+2$, owing to the fact that $N_\mu/T_s$ and $\mu/T_s$ are common to all $k$. Hence, it can be expected that the PAPR of LA$^2$FDM can also be significantly lower than that of AFDM.

Finally, for both the IA$^2$FDM and LA$^2$FDM, the signal $\bm{x}$ is formed by operating the corresponding inverse DA$^2$FT (IDA$^2$FT) on $\bm{s}$, yielding
\begin{equation}
\begin{aligned}
  \bm{x} &=\bm{A}^H\bm{s}= \bm{\Lambda}^H_{c_1}\bm{F}^H\bm{\Gamma}_\mu\bm{s}= \bm{\Lambda}^H_{c_1}\bm{F}^H\bm{P}\bm{\Upsilon}_\mu\bm{s}
\end{aligned}
\label{eq transmitted signal}
\end{equation}

\subsection{Input-Output Relationship}\label{subsection-A2FDM-proposed A2FDM Structure-I/O relationship}

Assume the double-selective fading channels having the CIR expressed
as \eqref{eq channel impulse response}. Then, the input-output
relationship in A$^2$FDM systems can also be represented by \eqref{eq
  demodulation} after the operation of DA$^2$FT at receiver,
where $\bm{H}_{eff}$ in \eqref{eq demodulation} can be expressed as
$\bm{H}_{eff}=\sum_{i=1}^Lh_i\bm{H}_i$ and
\begin{equation}
    \bm H_i \triangleq \bm A \bm \Gamma_{CPP_i}\bm \Delta_{fi}\bm \Pi^{\ell_i}\bm A^H.
    \label{eq Hi}
\end{equation}
Below we analyze the effective channel $\bm H_{eff}$ with the emphasis
on the effect from the matrix $\bm \Gamma_\mu$. We address here the
IA$^2$FDM, which can be readily generalized to the LA$^2$FDM.

After removing the CPP, it can be known that $\bm{\Gamma}_{CPP_i}$ becomes an identity matrix. In \eqref{eq Hi}, $\bm \Pi^{\ell_i}$ accounts for the delay of the $i$-th path, which has the element values given by $\delta(o-(n-\ell_i))$, where an element of $\bm \Pi^{\ell_i}$ is $1$ only if $o=n-\ell_i$ is satisfied, and otherwise, it is $0$. Note that, in $\delta(o-(n-\ell_i))$, $o$ and $n$ denote respectively the column index and row index of $\bm
\Pi^{\ell_i}$. Hence, with the above-mentioned limitations, the $(p,q)$-th element in $\bm H_i$ can be analyzed as
\begin{subequations}\label{eq element in Hi}
    \begin{align}
         H_i[p,q] &= \bm A[p,n]\bm \Delta_{f_i}[n,n]\bm \Pi^{\ell_i}[n,o]\bm A^H[o,q]\\
        &=\frac{1}{\sqrt{\mu}}\sum^{\mu-1}_{n=0}e^{-j2\pi(c_1n^2+\frac{pn}{N})}e^{-j2\pi f_in}\delta(o-(n-\ell_i))\frac{1}{\sqrt{\mu}}\sum^{\mu-1}_{o=0} e^{j2\pi(c_1o^2+\frac{qo}{N})}\\
        &=\frac{1}{\mu}\sum^{\mu-1}_{n=0}e^{-j2\pi(c_1n^2+\frac{pn}{N}+f_in)} e^{j2\pi\left(c_1n^2-2c_1\ell_in+c_1\ell_i^2+\frac{qn-q\ell_i}{N}\right)}\\
        &=\frac{1}{\mu}\sum^{\mu-1}_{n=0}e^{-j2\pi\left(\frac{(p-q)n}{N}+f_in+2c_1\ell_in-c_1\ell_i^2+\frac{q\ell_i}{N}\right)}\label{eq element in Hi4}\\
        &=\frac{1}{\mu}e^{j\frac{2\pi}{N}(Nc_1\ell_i^2-q\ell_i)}\sum^{\mu-1}_{n=0}e^{-j\frac{2\pi n}{N}(p-q+\nu_i+2Nc_1\ell_i)}\label{eq element in Hi5}\\
        &=\frac{1}{\mu}e^{j\frac{2\pi}{N}(Nc_1\ell_i^2-q\ell_i)}\mathcal{F}_i(p,q),\label{eq element in Hi6}
    \end{align}
\end{subequations}
where from \eqref{eq element in Hi4} to \eqref{eq element in Hi5} $\nu_i = Nf_i$ was applied followed by some arrangements afterwards. By definition, $\mathcal{F}_i(p,q)$ in \eqref{eq element in Hi6} is
\begin{equation}
\begin{aligned}
        \mathcal{F}_i(p,q)&=\sum^{\mu-1}_{n=0}e^{-j\frac{2\pi n}{N}(p-q+\nu_i+2Nc_1\ell_i)}\\
        &= \frac{1-e^{-j\frac{2\pi \mu}{N}(p-q+\nu_i+2Nc_1\ell_i)}}{1-e^{-j\frac{2\pi}{N}(p-q+\nu_i+2Nc_1\ell_i)}}\\
        &= \frac{1-e^{-j\frac{2\pi}{N_\mu}(p-q+\nu_i+2Nc_1\ell_i)}}{1-e^{-j\frac{2\pi}{N}(p-q+\nu_i+2Nc_1\ell_i)}},
\end{aligned}
\label{eq Fi(p,q)}
\end{equation}
which is dependent on the Doppler shift $\nu_i=\alpha_i+\beta_i$. As
seen, $\nu_i$ includes an integer part $\alpha_i$ and a fractional
part $\beta_i$, both imposing effect on $\mathcal{F}_i(p,q)$, as
analyzed below.

\subsubsection{Integer Doppler Shifts}\label{subsubsection-A2FDM-proposed A2FDM Structure-I/O relationship-integer doppler shift}

In this case, $\beta_i = 0$ and $\nu_i = \alpha_i$.  Define $\vartheta
=e^{-j\frac{2\pi}{N}(p-q+\nu_i+2Nc_1\ell_i)}$. Then, we can obtain
\begin{equation}
    \mathcal{F}_i(p,q) = 
    \begin{cases}
        \mu, &\vartheta=1,\\
        0, &\vartheta \neq 1.
    \end{cases}
\end{equation}
To get $\vartheta = 1$, $p-q+\alpha_i+2Nc_1\ell_i$ should be an integer that is a multiple of $N$, leading to $p =
[(\alpha_i+2Nc_1\ell_i-q)/N]$. Accordingly, \eqref{eq element in Hi} becomes
\begin{equation}
    H_i[p,q] = 
    \begin{cases}
        e^{j\frac{2\pi}{N}(Nc_1\ell_i^2-q\ell_i)}, &p = [(\eta_i-q)/N]\\
        0, &\text{otherwise},
    \end{cases}
    \label{eq Hi integer}
\end{equation}
where $\eta_i = [(\alpha_i+2Nc_1\ell_i)/N]$. Therefore, in row $p$ of $\bm H_i$, there is only one non-zero element, whose column number $q$ satisfies $p = [(\eta_i-q)/N]$. Consequently, $r(p)$ of the received signal $\bm{r}$ seen in the input-output relationship can be represented as
\begin{equation}
    r[p] = \sum^L_{i=1}h_ie^{j\frac{2\pi}{N}(Nc_1\ell_i^2-q\ell_i)}s[q]+\widetilde{w}[p], \quad p = 0,1,\cdots,N-1,
    \label{eq I/O relationship for integer doppler shift}
\end{equation}
where $q$ satisfies $p = [(\eta_i-q)/N]$.

\subsubsection{Fractional Doppler Shifts}\label{subsubsection-A2FDM-proposed A2FDM Structure-I/O relationship-fractional doppler shift}
In the case of fractional Doppler shifts, $-1/2<\beta_i \leq 1/2$ in $\nu_i = \alpha_i+\beta_i$. Hence, $\vartheta$ cannot be an integer value. Accordingly, \eqref{eq Fi(p,q)} can be simplified to an expression of
\begin{equation}
        \mathcal{F}_i(p,q) 
        = \frac{1-e^{-j2\pi \mu \theta}}{1-e^{-j2\pi \theta}}
        = \frac{\sin(\mu\phi)e^{-j (\mu-1) \phi}}{\sin\phi},
\end{equation}
where $\phi = \pi \vartheta = \frac{\pi}{N}(p-q+\nu_i+2Nc_1\ell_i)$ is defined. Then, the modulus of $H_i[p,q]$ can be simplified as
\begin{equation}
\begin{aligned}
        |H_i[p,q]| &= \left|\frac{1}{\mu}e^{j\frac{2\pi}{N}(Nc_1\ell_i^2-q\ell_i)}\mathcal{F}_i(p,q)\right|\\
        &= \left|\frac{1}{\mu}e^{j\frac{2\pi}{N}(Nc_1\ell_i^2-q\ell_i)}e^{-j (\mu-1) \phi}\frac{\sin(\mu\phi)}{\sin\phi}\right|\\
        &= \left|\frac{\sin(\mu\phi)}{\mu \sin\phi}\right|.
\end{aligned}
\end{equation}
Expressing $\mu\phi = (\mu-1)\phi+\phi$, $|H_i[p,q]|$ can be set to satisfy
\begin{equation}
    \begin{aligned}
        |H_i[p,q]|  &=\left|\frac{\sin((\mu-1)\phi+\phi)}{\mu \sin\phi}\right|\\
        & = \left|\frac{\sin((\mu-1)\phi)\cos\phi}{\mu \sin\phi}+\frac{\cos((\mu-1)\phi))}{\mu}\right|\\
        &\leq \left|\frac{\sin((\mu-1)\phi)\cos\phi}{\mu \sin\phi}\right|+\left|\frac{\cos((\mu-1)\phi))}{\mu}\right|.
    \end{aligned}
    \label{Hi[p,q] range analysis 1}
\end{equation}
Then, using the inequality $|\sin k\theta|\leq|k\sin\theta|$ for $k\geq 0$ and  $\cos((\mu-1)\phi)\leq 1$, we obtain
\begin{equation}
    \begin{aligned}
        |H_i[p,q]|
        &\leq \left|\frac{(\mu-1)\sin\phi \cos\phi}{\mu \sin\phi}\right|+\left|\frac{1}{\mu}\right|\\
        &\leq \frac{\mu-1}{\mu}|\cos\phi|+\frac{1}{\mu}\leq 1.
    \end{aligned}
    \label{Hi[p,q] range analysis 2}
\end{equation}

The right hand side of \eqref{Hi[p,q] range analysis 2} reaches its peak value $1$, when $\phi = v\pi$ for integer $v$. From $\phi =
\pi \theta = \frac{\pi}{N}(p-q+\nu_i+2Nc_1\ell_i)$, we know that the peak occurs at $q = [(p+\eta_i)/N]$, where $\eta_i =
[(\nu_i+2Nc_1\ell_i)/N]$ is defined. As $q$ moves away from this value, the right hand side of \eqref{Hi[p,q] range analysis 2} decreases. Moreover, the larger the value of $\mu$ is, the faster is the decreasing of $|H_i[p,q]|$ with respect to $q$. Therefore, following the analysis in \cite{bemani2023affine}, we approximate $|H_i[p,q]|$ to be non-zero only for $2\zeta + 1$ values of $q$, which are centered at $q = [(p+\eta_i)/N]$. Specifically, $\zeta$ can be set to a value so that, for all $i$ and $p$ satisfying $|q -
[(p+\eta_i)/N] |>\zeta$, the right hand side of \eqref{Hi[p,q] range analysis 2} is smaller than a pre-set small threshold. Accordingly, the elements in $\bm H_i$ have the values of
\begin{equation}
    H_i[p,q] = 
    \begin{cases}
        \frac{1}{\mu}e^{j\frac{2\pi}{N}(Nc_1\ell_i^2-q\ell_i)}\mathcal{F}_i(p,q), &\textrm{if}~[(p+{\eta}_i-\zeta)/N]\leq q \leq [(p+{\eta}_i+\zeta)/N],\\
        0, &\text{otherwise}.
    \end{cases}
    \label{eq Hi fractional}
\end{equation}

Consequently, in the case of fractional Doppler shifts, the input-output relationship of the A$^2$FDM can be represented as
\begin{equation}
\begin{aligned}
    r[p] = &\frac{1}{\mu}\sum^L_{i=1}\sum^{[(p+\eta_i+\zeta)/N]}_{q=[(p+\eta_i-\zeta)/N]}h_ie^{j\frac{2\pi}{N}(Nc_1\ell_i^2 -q\ell_i)}\\
    &\times\frac{1-e^{-j\frac{2\pi}{N_\mu}(p-q+\nu_i+2Nc_1\ell_i)}}{1-e^{-j\frac{2\pi}{N}(p-q+\nu_i+2Nc_1\ell_i)}}s[q]+\widetilde{w}[p]
\end{aligned}
    \label{eq I/O relationship for fractional doppler shift}
\end{equation}
for $p = 0,1,\cdots,N-1$.

\section{Analysis of the Effect of Parameters and Complexity}\label{section-A2FDM-parameter analysis}

In this section, we first analyze the effect of parameters $c_1$ and $\mu$ on diversity gain, followed by the analysis of parameter $\mu$ on the PAPR.

\subsection{Effect of Parameter $c_1$ on Diversity Order}\label{subsection-A2FDM-parameter analysis-c1 for diversity optimization}
For an AFDM having $N$ subcarriers and maximum Doppler shift $\nu_{\max}$ to achieve full diversity gain, it was suggested~\cite{bemani2023affine} that the parameter $c_1$ is set to $c_1 = \frac{2\nu_{\max}+1}{2N}$. This choice of $c_1$ is based on the observation that, to achieve full diversity gain, any distinct multipath components should make contributions via occupying non-overlapping positions in the effective channel matrices. Specifically, when two multipath components, $i$ and $j$, are considered, as shown in \cite{bemani2023affine}, the non-zero elements in $\bm H_i$ and $\bm H_j$ appear at the positions $[p, p +
  \eta_i]$ and $[p, p + \eta_j]$, respectively, were $p$ is the row index and $\eta_k \triangleq [(\nu_k + 2Nc_1\ell_k)/N]$, $k=i$ or $j$, with $[(\cdot)/N]$ denoting the modulo-$N$ operation. Then, to ensure the above-mentioned non-overlapping contributions made by components $i$ and $j$ in the effective channel matrix, the ranges covering $\eta_i$ and $\eta_j$ must be disjoint, i.e., satisfying
\begin{equation}
 \begin{aligned}
\{-\nu_{\max} + 2Nc_1\ell_i,\cdots,\nu_{\max} + 2Nc_1\ell_i\} \cap \\
 \{-\nu_{\max} + 2Nc_1\ell_j,\cdots,\nu_{\max} + 2Nc_1\ell_j\} = \emptyset,
 \end{aligned}
 \label{eq emptyest}
 \end{equation}
 where $\nu_{\max}$ is the maximum Doppler shift. Explicitly, this can  be achieved, if $c_1$ is chosen to satisfy \eqref{eq:optimum-c1}. If there are adjacent paths, meaning that $\min(|\ell_j - \ell_i|)=1$, $c_1$ is chosen to meet
 \begin{equation}
 c_1 > \frac{2\nu_{\max}}{2N}
 \label{eq c1 bound}
 \end{equation}
 which is typically selected as $c_1 = {(2\nu_{\max}+1)}/{2N}$.

The selection of $c_1$ based on \eqref{eq c1 bound} needs some channel knowledge of, at least $\nu_{\max}$, which may be time-varying in practice. For example, when an unmanned aerial vehicle (UAV) accelerates or turns around, the maximum Doppler shift might be different.  Based on \eqref{eq c1 bound}, we may argue that the condition can be satisfied, provided that a large value of $c_1$ is chosen to cover all the possible Doppler ranges.  However, \eqref{eq c1 bound} has been derived when the modulo-$N$ operation in $\eta_k$ is ignored.  If the modulo-$N$ operation is invoked, we will see that even when the condition of \eqref{eq c1 bound} is satisfied, two distinct multipath components may result in their contributions falling in the same location in the effective channel matrix, as shown below.

Since $\nu_i =f_iN=f_{Di}N/B$, where $f_{Di}$ is the actual Doppler
shift that is usually significantly smaller than $B$ of system's
bandwidth (or sampling rate), we can assume that $2\nu_{max}<N$. Now,
let us consider a $c_1$ that satisfies the condition of \eqref{eq c1
  bound}, but
 \begin{equation}\label{eq:A2FDM-52}
 \begin{aligned}
 \eta_i=\nu_i + 2Nc_1\ell_i &= \eta, \\
 \eta_j=\nu_j + 2Nc_1\ell_j &= dN + \eta,
 \end{aligned}
 \end{equation}
 where $0 \leq \eta \leq N-1$ and $d$ can be any non-zero
 integer. Then, although the indices of $\eta_i$ and $\eta_j$ are
 separated by a distance of $dN$, due to the modulo-$N$ operation, the
 contributions made by components $i$ and $j$ overlap in the effective
 channel matrix, hence, losing the diversity gain provided by
 multipath components $i$ and $j$. Furthermore, upon substituting
 $\eta$ from the first equation in \eqref{eq:A2FDM-52} into the second
 one, we can realize that, provided that there exist multipath
 components $i$ and $j$, whose associated quantities satisfy
 \begin{equation}
 c_1 = \frac{dN+\nu_i- \nu_j }{2N(\ell_j - \ell_i)},
 \label{eq overlap}
 \end{equation}
 one of the diversity orders then becomes lost. Therefore and in summary, the condition of \eqref{eq c1 bound} for $c_1$ selection is necessary but not sufficient.  Certain values of $c_1$ satisfying \eqref{eq c1 bound} may induce the overlaps of different multipath components in the effective channel matrix, resulting in the reduction of diversity gain, as demonstrated by the numerical results in Section~\ref{section-A2FDM-simulation results}.

Note that, according to the 3GPP's specifications \cite{3gpp38.901,3gpp38.211,3gpp38.913}, in most practical applications, Doppler shifts usually do not exceed $10\%$ of the subcarrier spacing, i.e., $\nu_i <0.1$. This means that the term of $\nu_i - \nu_j$ in \eqref{eq overlap} is negligible compared to $dN$, making \eqref{eq overlap} approximately $c_1 \approx {d}/{2(\ell_j -
  \ell_i)}$. When $c_1$ takes this value, it can be found that the non-zero elements in the effective channel matrix contributed by the $i$th and $j$th components are located at $[p, p+\nu_i]$ and $[p,
  p+\nu_j]$, respectively, which are effectively indistinguishable due to the small difference between $\nu_i$ and $\nu_j$. Moreover, if $c_1
\approx {d}/{2\min(|\ell_j - \ell_i|)}$, such as, $c_1 \approx
        {d}/{2}$ is used, all the multipath components will approximately satisfy the conditions in \eqref{eq:A2FDM-52}. In this case, although $c_1$ is chosen to satisfy \eqref{eq c1 bound}, the achieved diversity order will be reduced to approximately $1$, regardless of $L$, the actual number of multipaths.

In contrast, our proposed A$^2$FDM leverages both the chirp processing and precoding to achieve diversity gain. Specifically, each transmitted symbol is firstly spread onto $N_\mu$ subcarriers via the $\bm \Gamma_\mu$ operator. Hence, in A$^2$FDM systems with interleaved permutation, the achieved diversity order is $\min\{N_\mu, L\}$. Even when a $c_1$ is mistakenly chosen, leading to multipath component overlapping, the spreading by $\bm \Gamma_\mu$ operator is still capable of preserving the full achievable diversity gain, as long as $N_\mu \geq L$.  Otherwise, if $1 < N_\mu < L$, the achieved diversity order is still at least equal to $N_\mu$, instead of being receded to $1$ in the conventional AFDM systems. Note that, if $N_\mu = 1$, A$^2$FDM is reduced to the AFDM with $c_2 = 1$ or other integer.

In summary, while the diversity order achieved by AFDM may degrade to $1$ under the unfavourable setting of $c_1$, A$^2$FDM can consistently guarantee a diversity order of $\min\{N_\mu, L\}$, reducing the cost paid for a wrongly selected parameter $c_1$.  When $N_\mu \geq L$, the value of $c_1$ has no impact on the performance of A$^2$FDM, a full diversity gain can be guaranteed. If $N_\mu<L$, the full diversity gain can also be achieved, when $c_1$ is selected to meet the condition of \eqref{eq c1 bound} but not satisfy the condition in \eqref{eq overlap}.  Otherwise, if a wrong parameter $c_1$ is applied, the achieved diversity order is still guaranteed in the range of $N_\mu$ and $L$.

\subsection{Effect of Parameter $\mu$ on PAPR}\label{subsection-A2FDM-parameter analysis-mu for PAPR Reduction}
 AFDM inherits the same PAPR problem of OFDM, with the PAPR being as high as $N$ in an AFDM system with $N$ subcarriers.  Our A$^2$FDM scheme enables substantial mitigation of the PAPR problem conflicted by AFDM.

Specifically, for the IA$^2$FDM scheme addressed in Section \ref{subsubsection-A2FDM-proposed A2FDM Structure-design of DA2FT-IDA2FT}, as shown in \eqref{eq:A2FDM-29}, there are only $\mu$ out of the $N$ subcarriers activated at any time, with the $\mu$ subcarriers uniformly distributed within the $N$ subcarriers. Similarly, for the LA$^2$FDM considered in Section \ref{subsubsection-A2FDM-proposed A2FDM Structure-design of DA2FT-LDA2FT}, as illustrated by \eqref{eq:A2FDM-35}, there are either $\mu$ or $(\mu+2)$ out of the $N$ subcarriers activated, depending on the time instants within a symbol duration. Hence, in both IA$^2$FDM and LA$^2$FDM, the parameter $\mu$ can be adjusted to strike a trade-off between PAPR and implementation complexity, as analyzed in the following section. For instance, if $\mu=1$, IA$^2$FDM activates only one subcarrier while LA$^2$FDM activates three, regardless of the value of $N$. However, in this case, A$^2$FDM needs to add most computation on AFDM. By contrast, when $\mu$ increases, signal's PAPR in A$^2$FDM increases, but the extra computation invested above AFDM reduces. Nevertheless, as the extra computation is mainly from the blocks of FFT operations, the complexity of A$^2$FDM added on AFDM is mild, as to be detailed in Section~\ref{subsection-A2FDM-Complexity-Comparison}.

\subsection{Complexity Comparison between A$^2$FDM and AFDM}\label{subsection-A2FDM-Complexity-Comparison}

At transmitter, by comparing the $\bm{A}$ matrices in AFDM and in
A$^2$FDM, which are $\bm{A}=\bm{\Lambda}_{c_2}\bm{F}
\bm{\Lambda}_{c_1}$ for AFDM and
$\bm{A}=\bm{\Upsilon}_\mu^H\bm{P}^T\bm{F}\bm{\Lambda}_{c_1}$ for
A$^2$FDM, we can readily know the extra computation required by
A$^2$FDM in comparison with AFDM. As shown in these formulas, to
implement A$^2$FDM, the matrix $\bm{\Lambda}_{c_2}$ is replaced by
$\bm{\Upsilon}_\mu^H\bm{P}^T$. For AFDM transmitter, the complexity
due to $\bm{\Lambda}_{c_1}$ and $\bm{\Lambda}_{c_2}$ can be ignored,
making the complexity of AFDM is similar to that of OFDM, which is
$\mathcal{O}(N\ln N)$, contributed by the $N$-point IFFT.  For
A$^2$FDM transmitter, the complexity contributed by
$\bm{\Lambda}_{c_1}$ and $\bm{P}$ of permutation operation can also be
ignored. Then, the complexity of A$^2$FDM transmitter is given by that
for computing the $\mu$ number of $N_\mu$-point FFT and that for
computing one $N$-point IFFT, yielding a complexity of
$\mathcal{O}(N[\ln N+\ln (N/\mu)])$. Hence, the extra complexity of
A$^2$FDM transmitter above AFDM transmitter is $\mathcal{O}(N\ln
(N/\mu))$, which reduces as $\mu$ increases\footnote{Note that, more
accurately, A$^2$FDM transmitter also has the complexity of
$\mathcal{O}(N\ln N)$ as $\mathcal{O}(N[\ln N+\ln
  (N/\mu)])=c\mathcal{O}(N\ln N)$ considering $\mu<N$, where $c$ is a
constant. Hence, in comparison with AFDM, A$^2$FDM only has some extra
computation that is proportional to $\mathcal{O}(N\ln
(N/\mu))$}. However, as discussed in
Section~\ref{subsection-A2FDM-parameter analysis-mu for PAPR
  Reduction}, the PAPR of A$^2$FDM signals increases, as $\mu$
increases, making a trade-off between the complexity of transmitter
implementation and the PAPR of transmit signals.

At receiver, both AFDM and A$^2$FDM require the equalization built on the signals received from $N$ subcarriers. Accordingly, the inverse of a $(N\times N)$ matrix is required, for instance, if a MMSE-based equalizer is implemented. This equalizer dominates receiver's computation, having the complexity much higher than the $N$-point FFT. Hence, the computational loads for both AFDM and A$^2$FDM receivers are similar.

\section{Performance Results and Analysis}\label{section-A2FDM-simulation results}

In this section, we present the simulation results to demonstrate the
performance of our proposed A$^2$FDM and compare it with the
performance of AFDM. The main parameters used in simulations are the
same as listed in \tref{tab:system parameters}, unless otherswise
specifid. The delays of multipath components normalized by the
sampling duration are randomly distributed in the integer set
$[1,\ell_{\max}=30]$, where $\ell_{\max}$ denotes the maximum
normalized delay-spread. The Doppler shifts normalized by subcarrier
spacing $\Delta f$ is distributed in $[-\nu_{\max},\nu_{\max}]$, set
according to $\nu_l=\nu_{\max}\text{cos}(\varphi_i)$ with $\varphi_i$
uniformly distributed in $[-\pi, \pi]$. In the cases of achieving
maximum diversity order, the value of parameter $c_1$ is typically set
as $c_1 = c_{1f}$, as seen in \eqref{eq:optimum-c1}, for both AFDM and
A$^2$FDMA, unless otherwise specified.  For receiving equalization,
the MMSE-assisted detection is employed.

\begin{figure}[htbp]
    \centering
    \includegraphics[width = 0.65\linewidth]{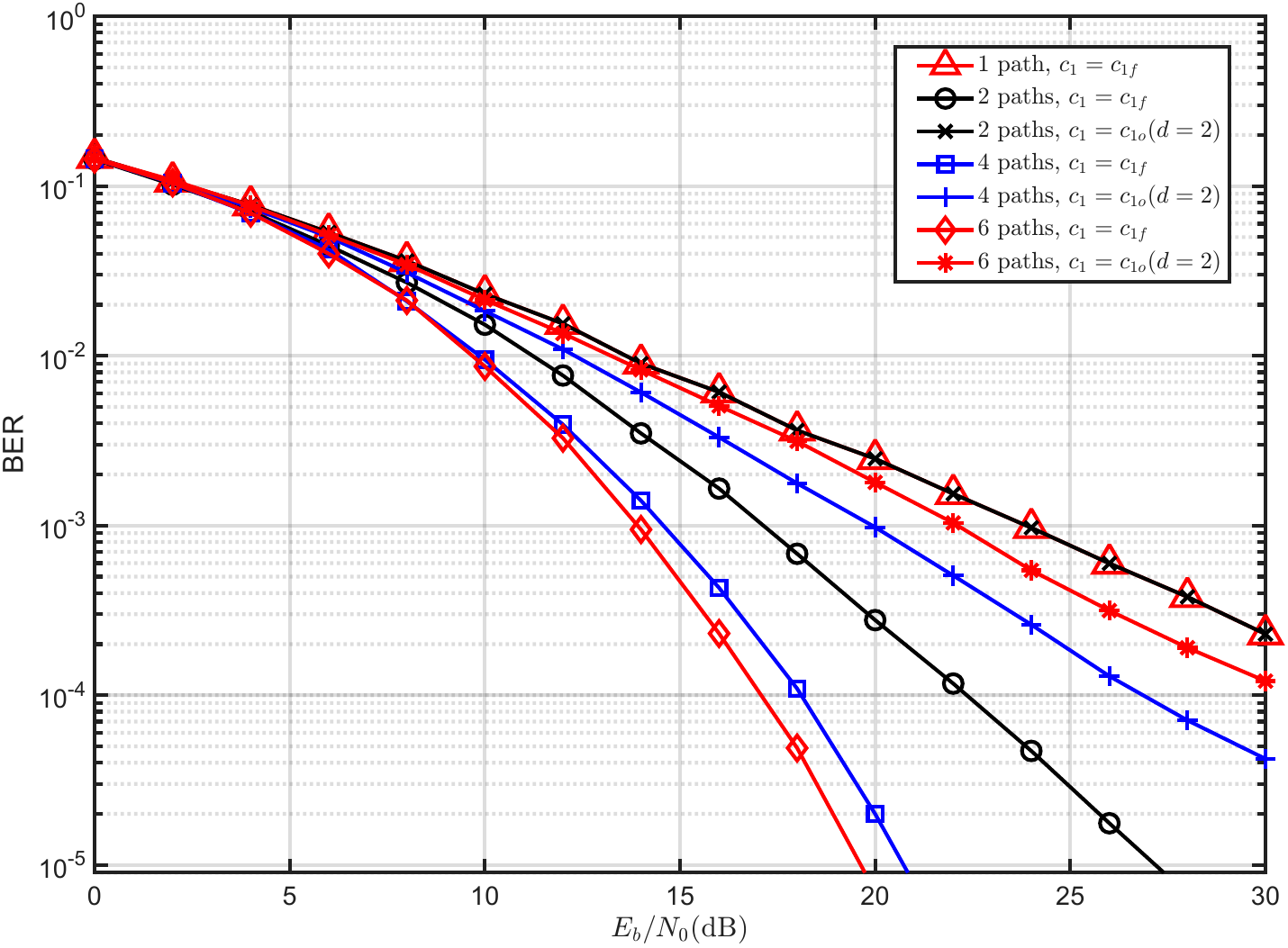}
    \caption{Demonstration of the effect of $c_1$ on the BER performance of AFDM systems.}
    \label{fig:conventional AFDM-pa}
\end{figure}

In Section~\ref{subsection-A2FDM-parameter analysis-c1 for diversity
  optimization}, the analysis shows that an inappropriate value for
$c_1$ may result in the loss of performance of the AFDM systems,
although $c_1$ is set to satisfy \eqref{eq:optimum-c1} but also
satisfies \eqref{eq overlap}. This is shown in
Fig.~\ref{fig:AFDM-with-different-pb} in
Section~\ref{subsection-A2FDM-conventional AFDM
  structure-Demodulation} and Fig.~\ref{fig:conventional AFDM-pa}
here, when $c_1$ is set according to \eqref{eq overlap} with $d=2$,
labelled as $c_1=c_{1o}(d=2)$ in these figures. As shown in
Fig.~\ref{fig:AFDM-with-different-pb}, most of the diversity gain is
lost, although wireless channel has $L=10$ paths. As shown in
Fig.~\ref{fig:conventional AFDM-pa}, the BER performance becomes
unpredictable, with $L=2$ yielding nearly the same performance as
$L=1$, and the BER performance of $L=4$ is better than that of
$L=6$. In all these cases, the BER performance is much worse than that
of the corresponding cases with $c_1 = c_{1f}$.

\begin{figure}[htbp]
    \centering
    \includegraphics[width = 0.65\linewidth]{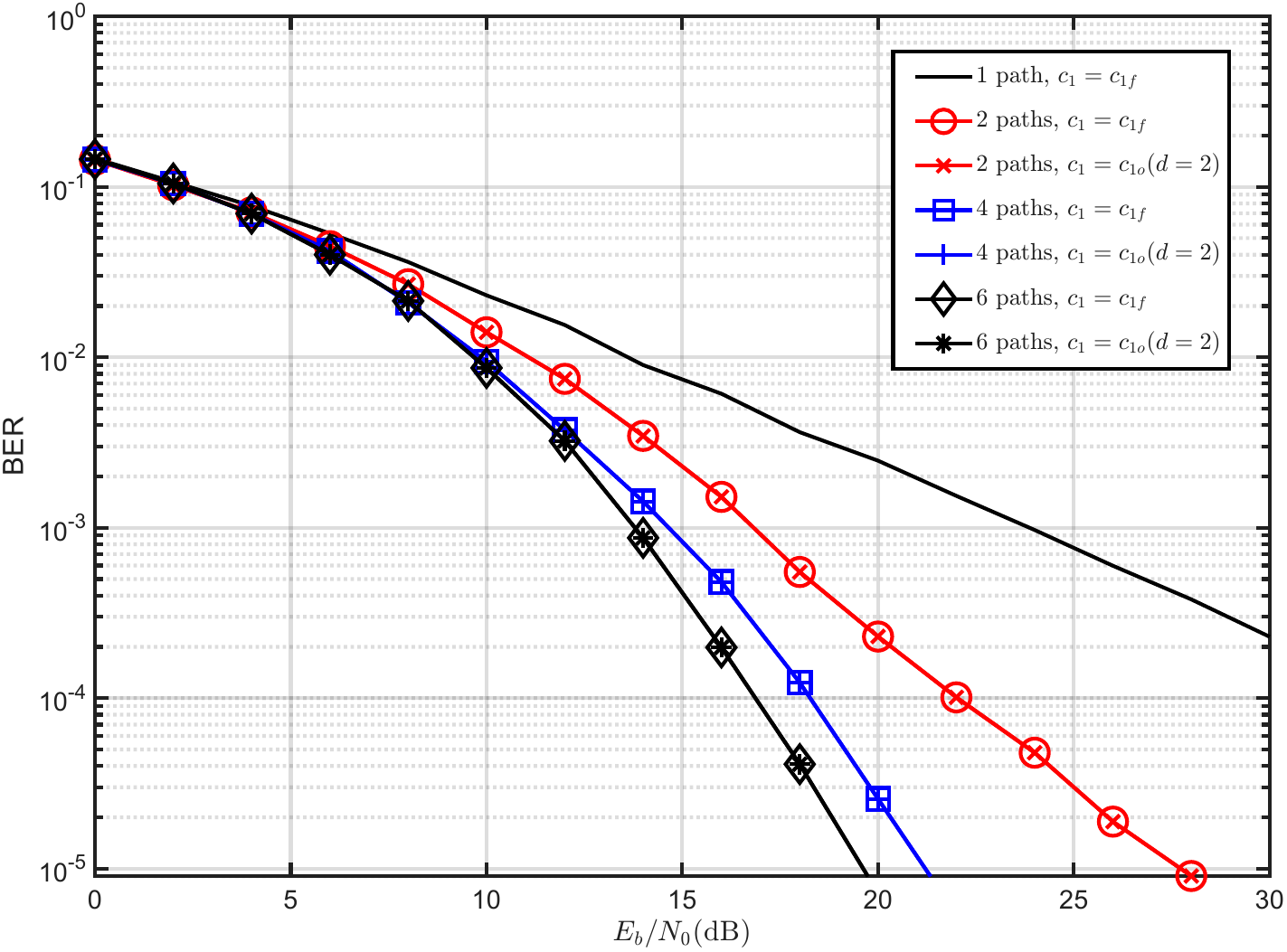}
    \caption{Demonstration of the effect of $c_1$ on the BER performance of IA$^2$FDM systems with $\mu$ set to a relatively large value of $\mu=128$. }
    \label{fig:conventional AFDM-pc}
\end{figure}

\begin{figure}[htbp]
    \centering
    \includegraphics[width = 0.65\linewidth]{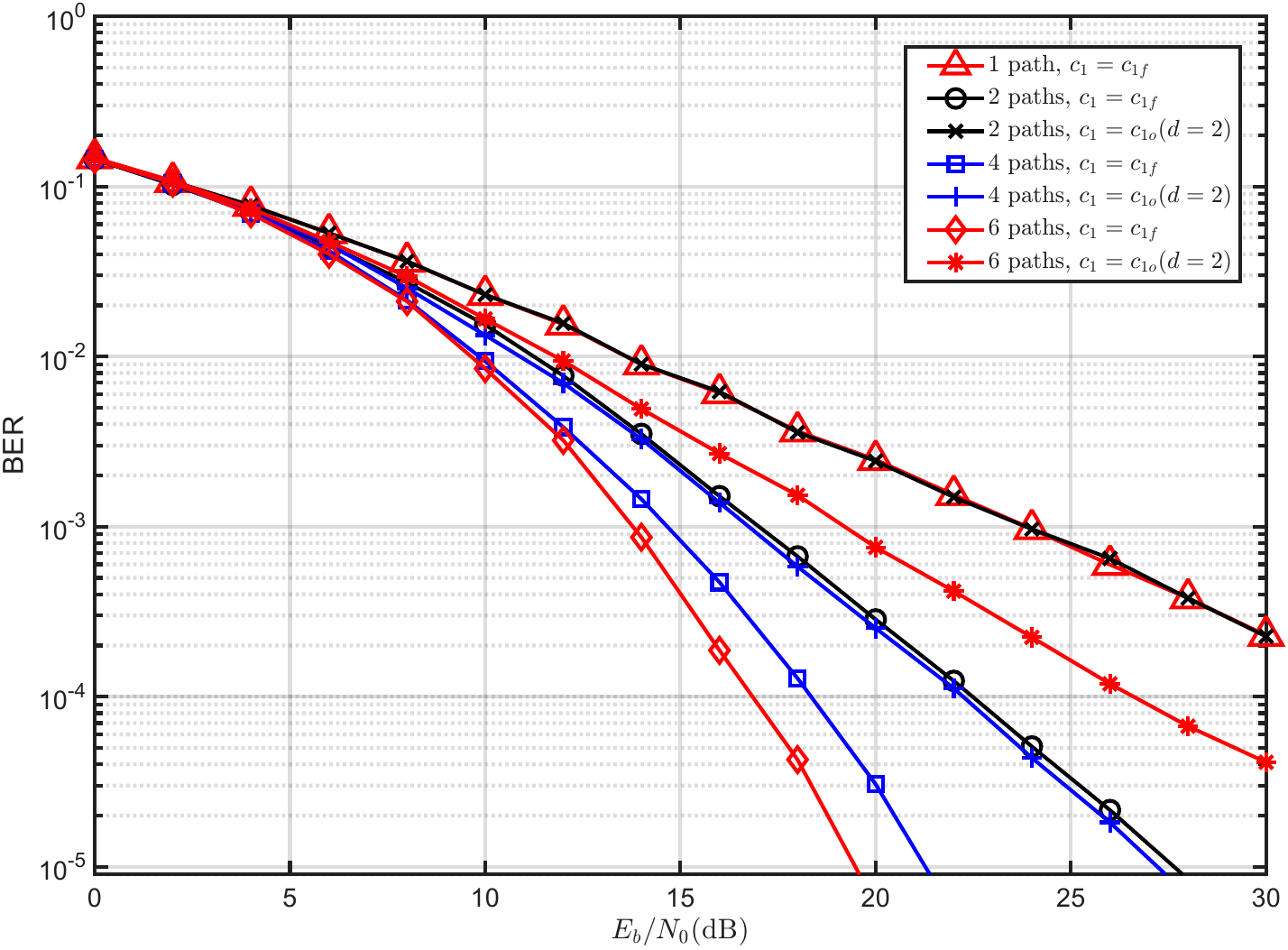}
    \caption{Demonstration of the effect of $c_1$ on the BER performance of IA$^2$FDM systems with $\mu$ set to a relatively small value of $\mu=2$.}
    \label{fig:conventional AFDM-pe}
\end{figure}

Correspondingly, in Figs.~\ref{fig:conventional AFDM-pc} and
\ref{fig:conventional AFDM-pe}, the BER performance of the IA$^2$FDM
systems with relatively large $\mu=128$ and small $\mu=2$ is
demonstrated. As seen in Fig.~\ref{fig:conventional AFDM-pc}, for a
given $L$ of the number of paths, the BER performance is the same in
the cases of employing both appropriate and inappropriate $c_1$
values. This is because the achieved diversity gain in the case of
relatively large $\mu=128$ is mainly determined by $\mu$. By contrast,
when $\mu=2$, which is small, as shown in Fig.~\ref{fig:conventional
  AFDM-pe}, the BER performance becomes similar as that shown in
Fig.~\ref{fig:conventional AFDM-pa}. However, when comparing the
results shown in Fig.~\ref{fig:conventional AFDM-pe} and that in
Fig.~\ref{fig:conventional AFDM-pa}, if the value of $c_1$ is
correctly set, both AFDM and IA$^2$FDM attain the same BER
performance. In contrast, when $c_1$ is wrongly set, the BER
performance of IA$^2$FDM is noticeably better than that of AFDM for
both $L=4$ and $L=6$, in addition to the reduced PAPR in IA$^2$FDM
systems.

\begin{figure}[htbp]
    \centering
    \includegraphics[width=0.65\linewidth]{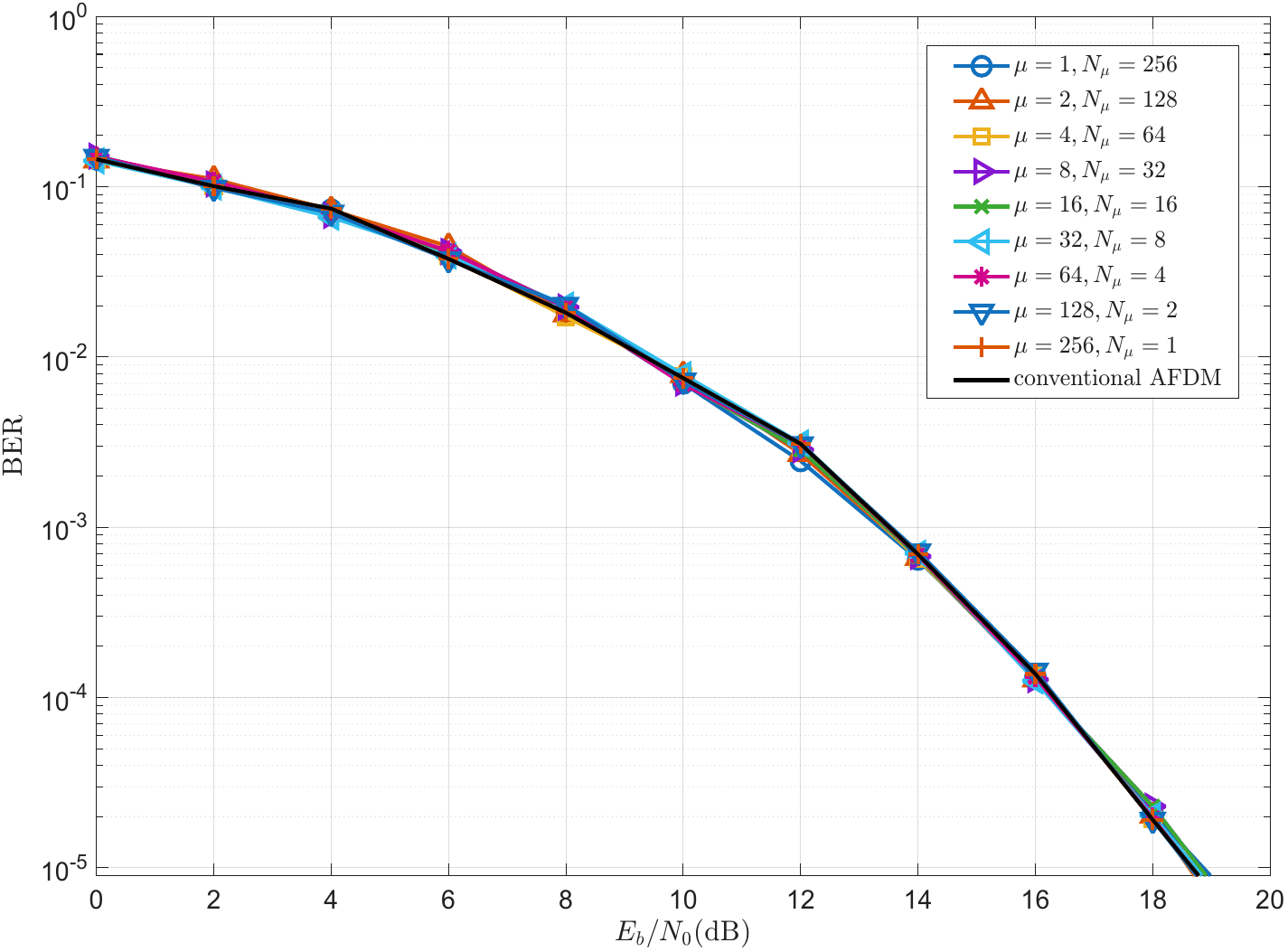}
    \caption{BER performance of IA$^2$FDM with $\mu=1,2,4,8,16,32,64,128$ and $256$, when $c_1=c_{1f}$ is set.}
    \label{fig:BER IDA2FT}
\end{figure}

\fref{fig:BER IDA2FT} illustrates the BER performance of the IA$^2$FDM
systems with configurations of $\mu=1, 2, 4, 8, 16, 32, 64, 128$ and
$256$, respectively. Accordingly, $N_\mu$ has the values of
$N_\mu=256$, $128$, $64$, $32$, $16$, $8$, $4$, $2$, and $1$,
respectively. As the baseline, the BER performance of the
corresponding AFDM system is included. As shown in \fref{fig:BER
  IDA2FT}, the BER performance of IA$^2$FDM systems is nearly the same
as that of the AFDM scheme. When $\mu=1$, from \eqref{Block DAFT} in
Section~\ref{subsection-A2FDM-proposed A2FDM Structure-design of
  DA2FT}, we can know that $\bm{A}=\bm{\Lambda}_{c_1}$, and hence,
IA$^2$FDM is reduced to a single-carrier signalling scheme. This
single-carrier system is expected to experience severer ISI. However,
unlike the conventional single-carrier systems, the employment of CPP
in IA$^2$FDM effectively mitigates the ISI caused by the
single-carrier signalling, enabling the system to achieve the
comparable performance under both single-carrier and multi-carrier
configurations.

\begin{figure}[htbp]
    \centering
    \includegraphics[width=0.65\linewidth]{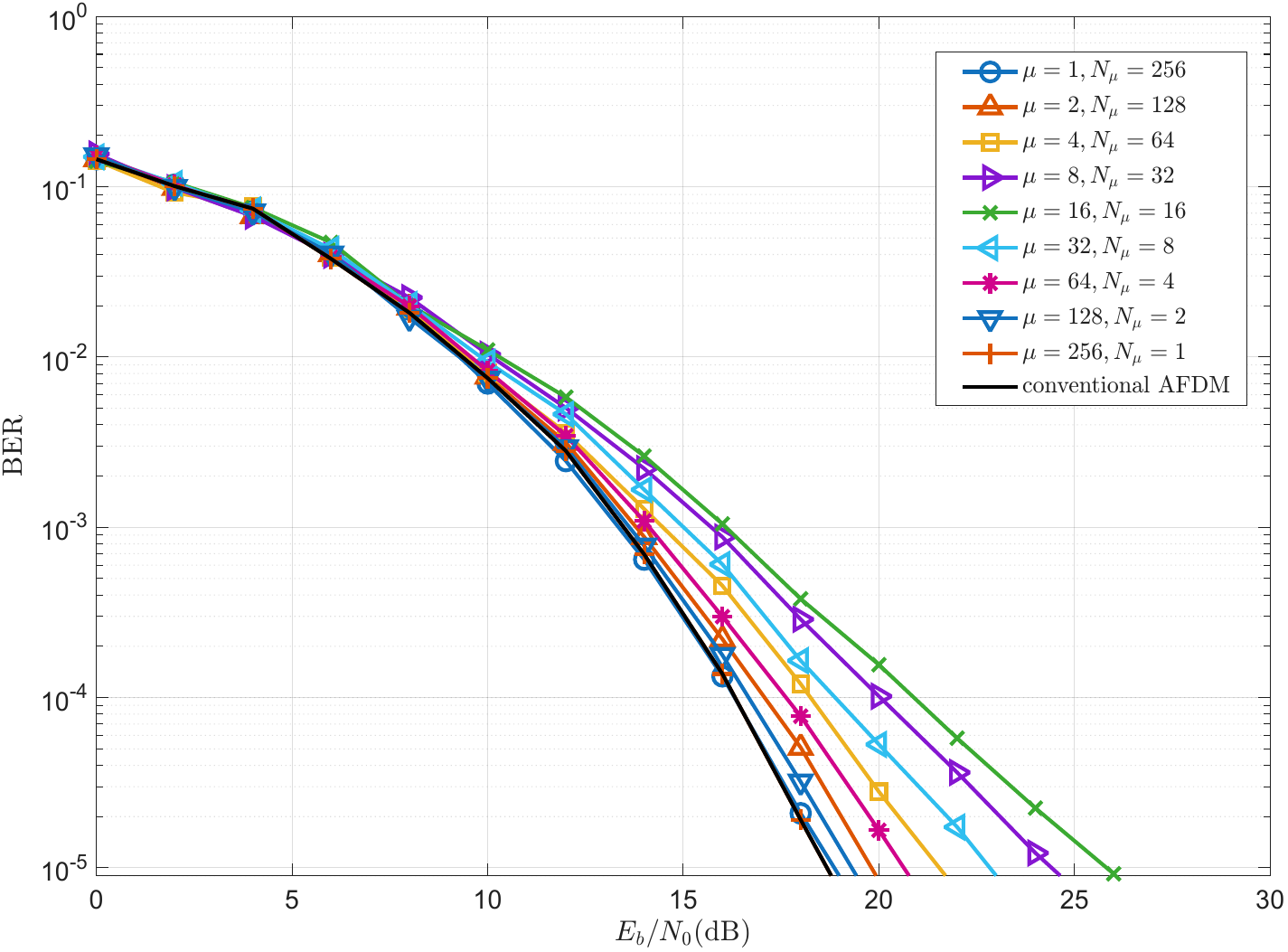}
    \caption{BER performance of LA$^2$FDM with $\mu=1,2,4,8,16,32,64,128$ and $256$, when $c_1=c_{1f}$ is used.}
    \label{fig:BER LDA2FT}
\end{figure}

\fref{fig:BER LDA2FT} shows the BER performances of the LA$^2$FDM
systems with $\mu = 1$, $2$, $4$, $16$, $32$, $64$, $128$ and $256$,
respectively, yielding corresponding $N_\mu=256$, $128$, $64$, $32$,
$16$, $8$, $4$, $2$, and $1$. Again, the BER performance of the AFDM
attaining full diversity gain is provided as the benchmark. As shown
by the results, the BER performance of LA$^2$FDM first deteriorates as
$\mu$ increases from $\mu=1$ to $\mu = 16$, and then the BER
performance improves as $\mu$ further increases, ultimately converging
to that of the AFDM obtaining full diversity gain.  The reason for the
performance of LA$^2$FDM systems as shown in \fref{fig:BER LDA2FT} is
as follows. First, with LA$^2$FDM, a sub-block of data symbols,
$\bm{s}_i$, after FFT processing is assigned to adjacent subcarriers,
which experience correlated fading. The diversity gain contributed by
the operator $\bm{\Gamma}_{\mu}$ in LA$^2$FDM is usually only
attainable when the number of subcarriers per block, $N_\mu$, is
larger than $L$, a larger $N_\mu$ yields a higher diversity
gain. Second, when $\mu$ is smaller, LA$^2$FDM becomes more like
single-carrier, yielding higher ISI, which degrades the BER
performance of LA$^2$FDM systems. Third, as $\mu$ increases or $N_\mu$
decreases, LA$^2$FDM converges to AFDM, yielding reduced ISI with
improved diversity gain due to the operation of
$\bm{\Lambda}_{c_1}$. After combining all the above-mentioned effects,
the BER performance of LA$^2$FDM systems first degrades, as $\mu$
increases, which is due to the reduced diversity gain contributed by
the $\bm{\Gamma}_{\mu}$ related operations, and then improves, as
$\mu$ further increases, which is contributed by the diversity gain
provided by the $\bm{\Lambda}_{c_1}$ relied operation.

Note that in LA$^2$FDM, since the sub-blocks of data symbols are
mapped to adjacent subcarriers experiencing correlated fading, the
$\bm{\Gamma}_{\mu}$ related operations, i.e., FFT and mapping, may not
introduce diversity gain. However, this transmission scheme is
beneficial to the downlink resource-allocation, where a sub-block of
adjacent subcarriers, such as $N_\mu$ subcarriers, can be assigned to
one downlink user. This downlink transmission scheme is usually
capable of obtaining multiuser diversity gain, as different downlink
users at different locations are expected to experience independent
fading.

\begin{figure}[th]
    \centering
    \includegraphics[width=0.65\linewidth]{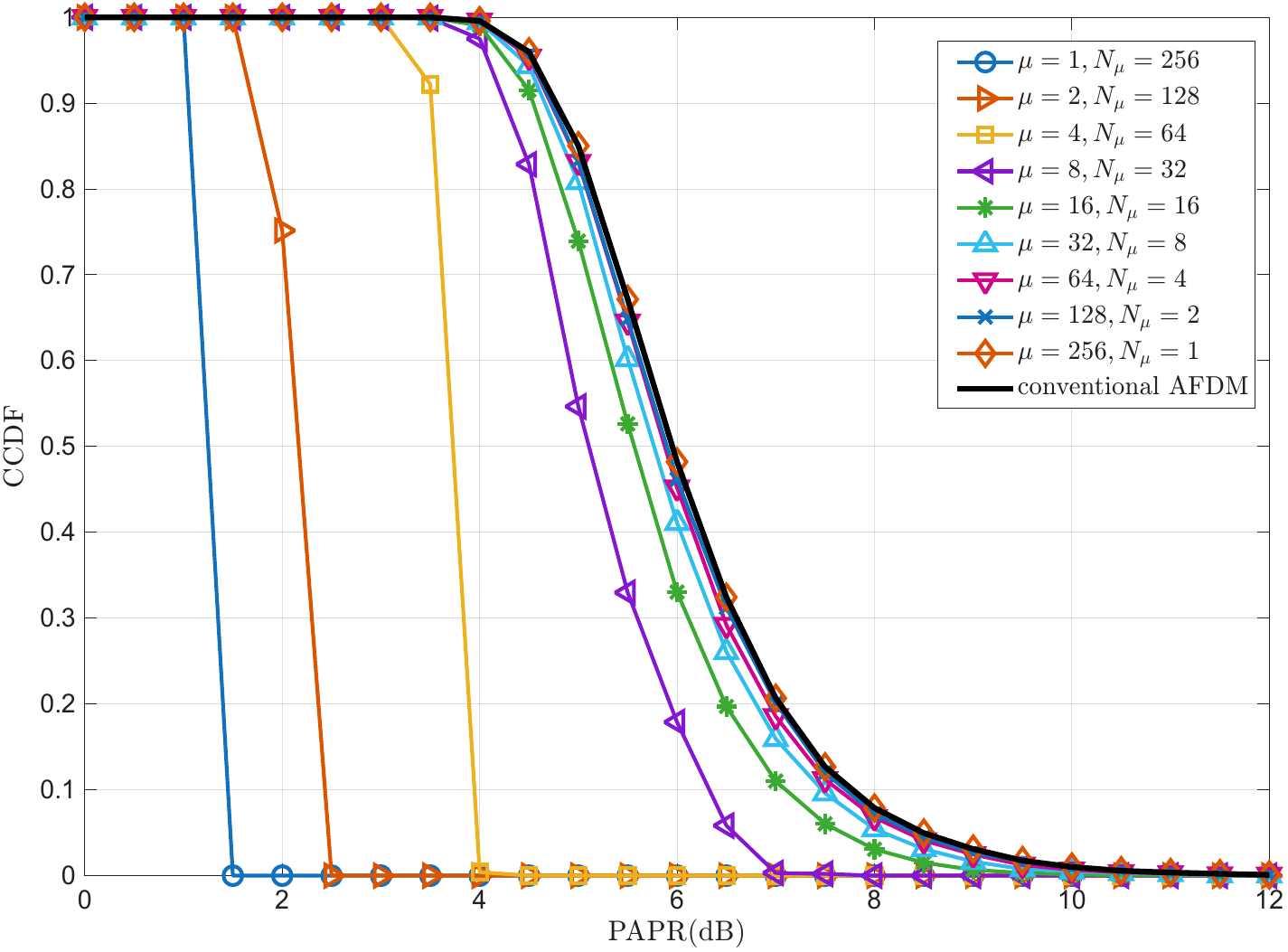}
    \caption{CCDF of PAPR of the IA$^2$FDM with $\mu=1,2,4,8,16,32,64,128$ and $256$.} 
    \label{fig:PAPR IDA2FT}
\end{figure}
\begin{figure}[th]
    \centering
    \includegraphics[width=0.65\linewidth]{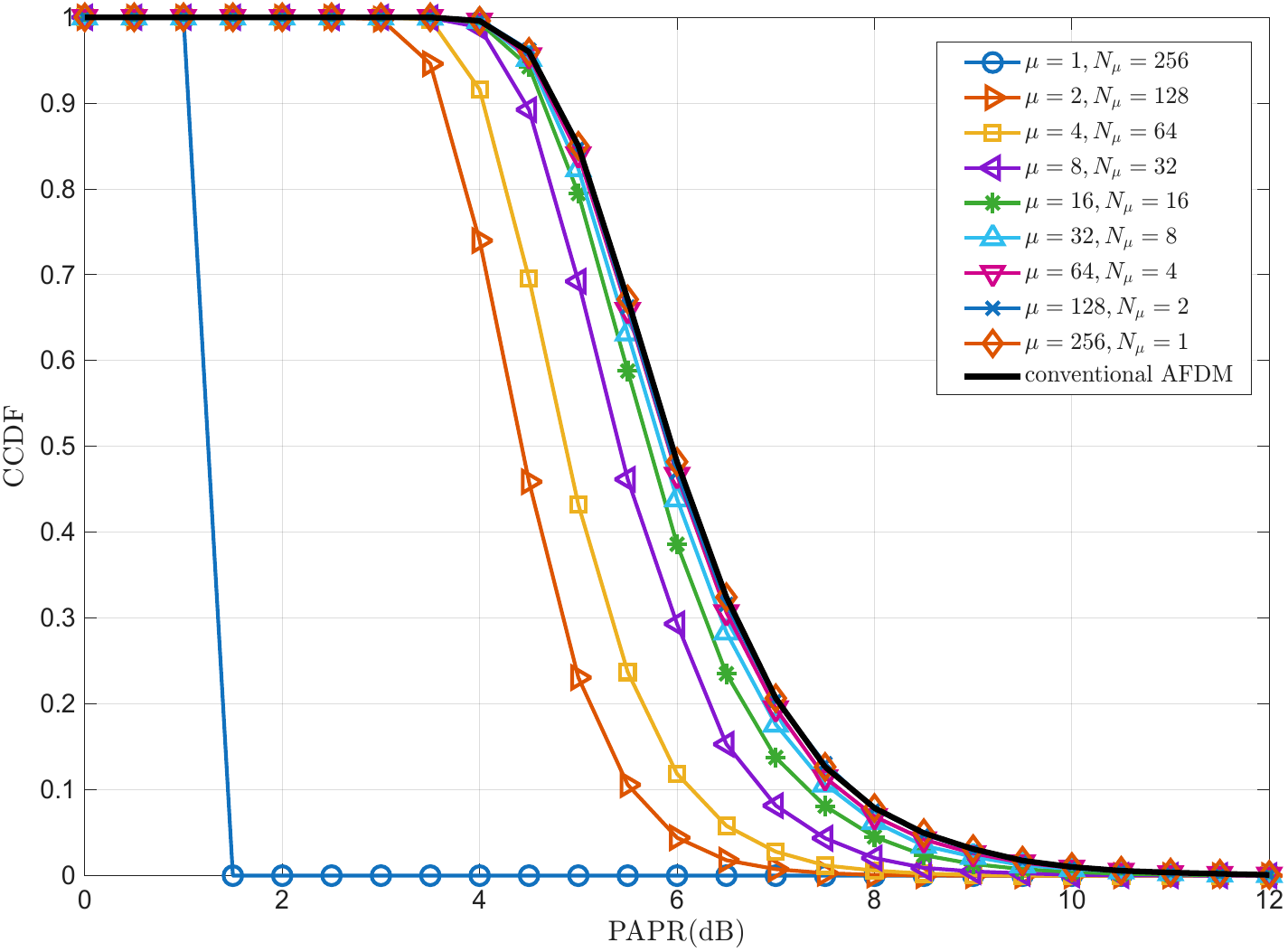}
    \caption{CCDF of PAPR of the LA$^2$FDM with $\mu=1,2,4,8,16,32,64,128$ and $256$.}
    \label{fig:PAPR LDA2FT}
\end{figure}

Figs.~\ref{fig:PAPR IDA2FT} and \ref{fig:PAPR LDA2FT} depict
respectively the Complementary Cumulative Distribution Functions
(CCDFs) of PAPR of the IA$^2$FDM and LA$^2$FDM for $\mu=1, 2, 4, 8,
16, 32, 64, 128$ and $256$. For comparison, the CCDF of PAPR of the
AFDM is included. When $\mu=1$, both the IA$^2$FDM and LA$^2$FDM
reduce to the single-carrier scheme, hence having the lowest
PAPR. Otherwise, according to the analysis in Sections
\ref{subsubsection-A2FDM-proposed A2FDM Structure-design of
  DA2FT-IDA2FT} and \ref{subsubsection-A2FDM-proposed A2FDM
  Structure-design of DA2FT-LDA2FT}, the PAPR of both IA$^2$FDM and
LA$^2$FDM increases with the increases $\mu$, as explicitly seen in
both figures, when $\mu$ is relatively small.  By contrast, when the
number of subcarriers is relatively big, which is $256$ for the two
figures, PAPR can be closely approximated by Gaussian
distribution. Consequently, if $\mu$ is big enough, the PAPR's CCDFs
of both the IA$^2$FDM and LA$^2$FDM converge to that of the
AFDM. However, we should note that although the distribution of PAPR
of the AFDM becomes similar as that of the IA$^2$FDM and LA$^2$FDM,
the worst PAPR of AFDM is still $N$, the number of subcarriers. In
contrast, the worst PAPR of, for example, the IA$^2$FDM is only $\mu$.

When comparing \fref{fig:PAPR IDA2FT} and \fref{fig:PAPR LDA2FT}, we
can see that, for a given $\mu$ - especially a given small value of
$\mu$, the PAPR of LA$^2$FDM is slightly higher than that of
IA$^2$FDM, which agrees with our analysis in Sections
\ref{subsubsection-A2FDM-proposed A2FDM Structure-design of
  DA2FT-IDA2FT} and \ref{subsubsection-A2FDM-proposed A2FDM
  Structure-design of DA2FT-LDA2FT}.

\begin{figure}[htbp]
    \centering
    \includegraphics[width=0.65\linewidth]{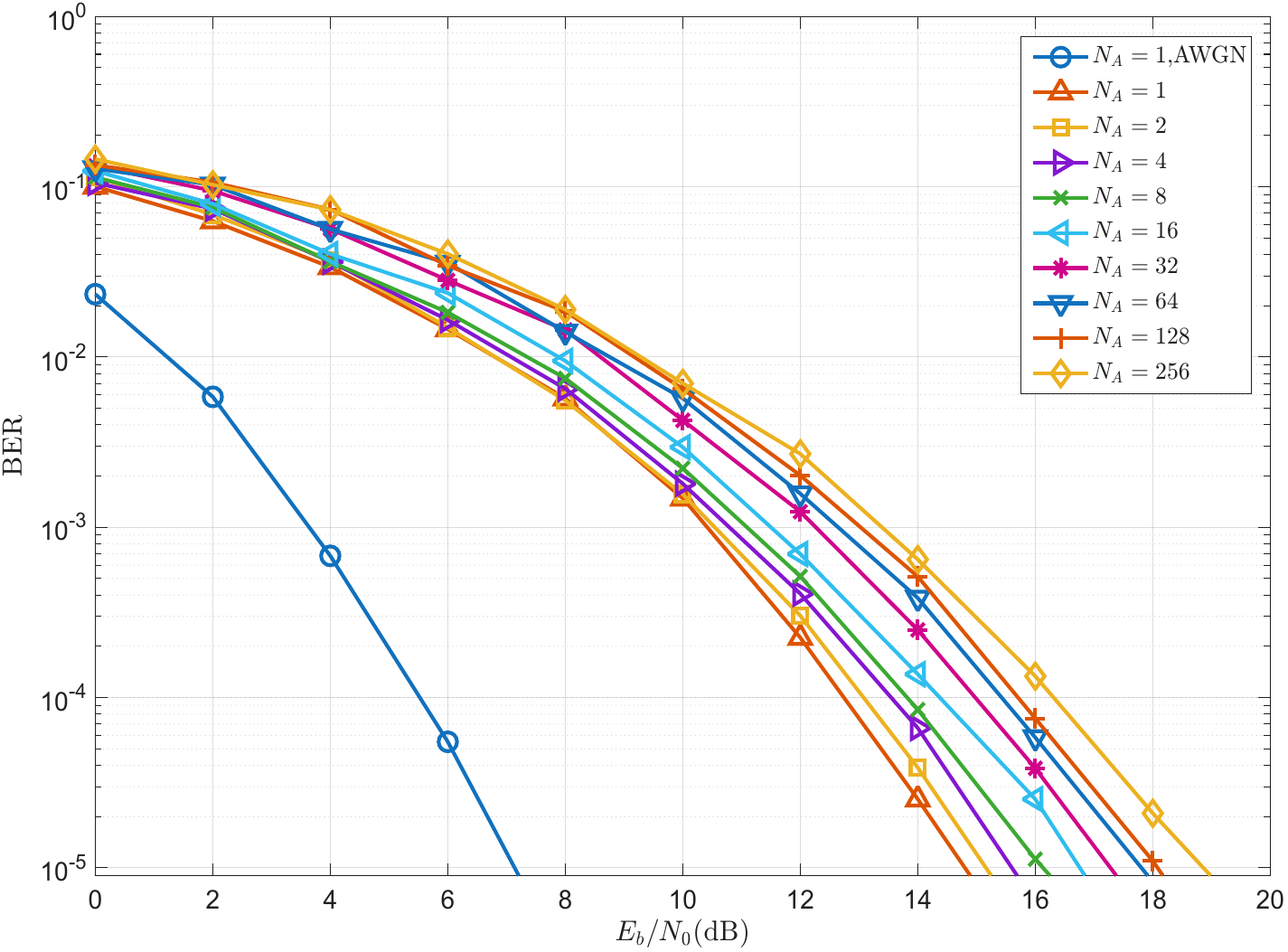}
    \caption{Data rate and BER trade-off of the IA$^2$FDM systems with
      $N_A$ out of $N$ symbols activated for transmission, when
      $\mu=1$ is assumed.}
    \label{fig:different activated symbol mu1}
\end{figure}
\begin{figure}[htbp]
    \centering
    \includegraphics[width=0.65\linewidth]{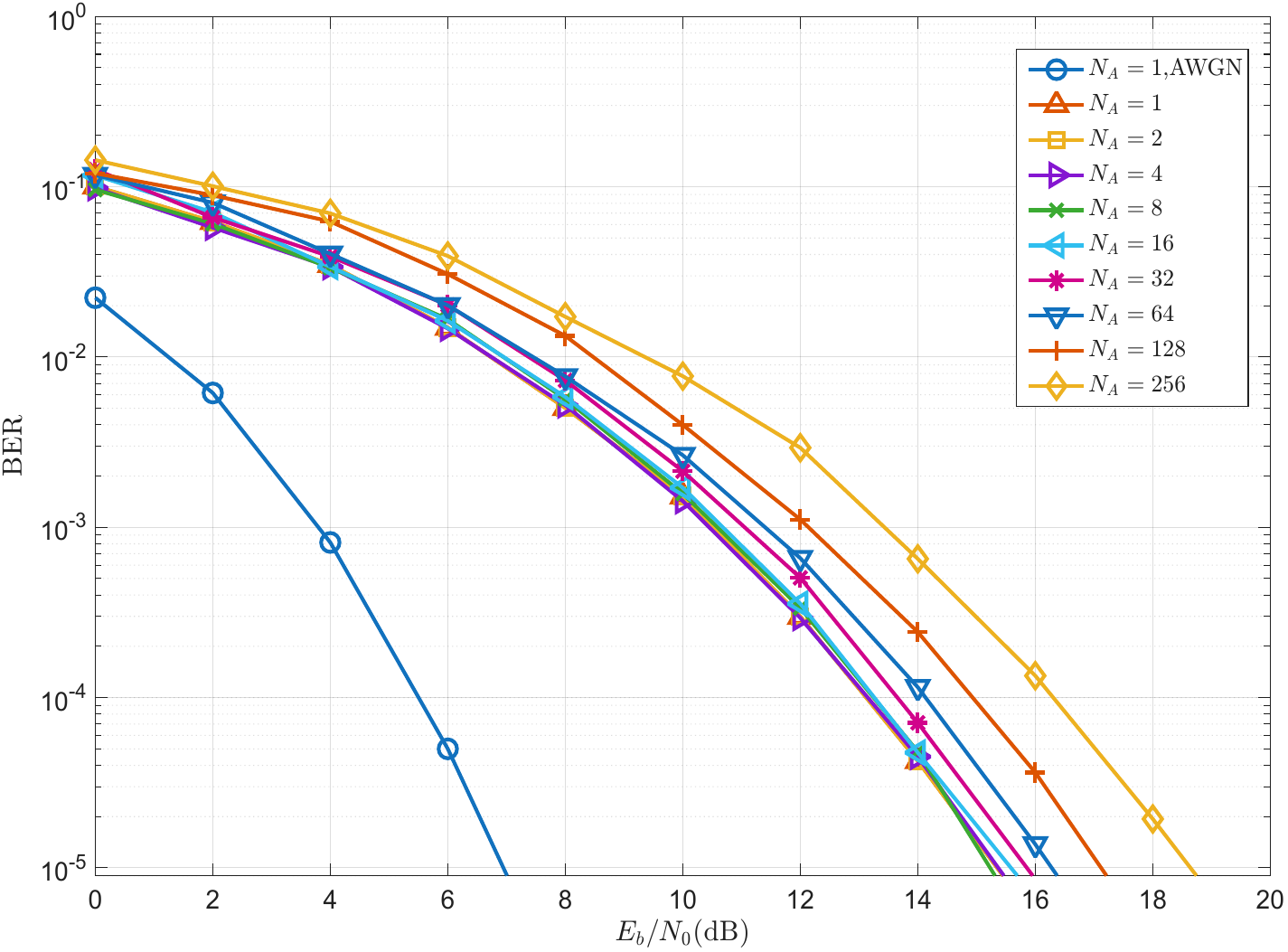}
    \caption{Data rate and BER trade-off of the IA$^2$FDM systems with
      $N_A$ out of $N$ symbols activated for transmission, when
      $\mu=4$ is assumed.}
    \label{fig:different activated symbol mu4}
\end{figure}

Finally, in Figs.~\ref{fig:different activated symbol mu1} and
\ref{fig:different activated symbol mu4} , the trade-off between data
rate and diversity gain is demonstrated, when assuming $\mu=1$ and
$\mu=4$, respectively, where the BER performance of 4QAM over Gaussian
channel is also included for reference. Note that in these two
figures, $N_A$ From the principles of IA$^2$FDM, and also of AFDM, we
know that it experiences ISI, which needs to be sufficiently
suppressed to obtain diversity gain. Higher ISI needs more resources,
or more signal processing effort, to suppress, which may result in
less diversity gain available for signal detection. In this example,
activating more symbols generates higher ISI. Hence, when $N_A=1$,
there is no ISI. Hence, full diversity gain provided by the
$L=10$-path double-selective fading channel can be obtained. Then, as
$N_A$ increases, which yields more ISI, the BER performance degrades,
explicitly showing the trade-off between data rate and BER
performance, or diversity gain.

When comparing Figs.~\ref{fig:different activated symbol mu1} and
\ref{fig:different activated symbol mu4}, we can find that, except the
cases of $N_A=1,~2$, the IA$^2$FDM with $\mu=4$ provides better BER
performance than the IA$^2$FDM with $\mu=1$ for all the other cases
with $N_A>2$. The reason behind is that the IA$^2$FDM with $\mu=1$ is
a single-carrier scheme, while the IA$^2$FDM with $\mu=4$ uses $4$
subcarriers to transmit at every time instant. Accordingly, in the
IA$^2$FDM with $\mu=1$, all symbols are transmitted on one subcarrier
at any time instant, while in the IA$^2$FDM with $\mu=4$, only
one-fourth of symbols are transmitted on each of the four active
subcarriers at any time instant. Consequently, when $N_A=1$ or $2$,
meaning that there is no ISI or small ISI, the IA$^2$FDM with $\mu=1$
does not need to make much effort for ISI suppression and hence, can
enjoy the near-full diversity gain. Otherwise, when $N_A>2$, the
IA$^2$FDM with $\mu=1$ is required to make more effort to suppress ISI
than the IA$^2$FDM with $\mu=4$, which experiences less ISI owing to
using $4$ subcarriers at any time instant. Consequently, the IA$^2$FDM
with $\mu=1$ attains reduced diversity gain, making its BER
performance become worse than that of the IA$^2$FDM with $\mu=4$.

\section{Conclusions}\label{section-A2FDM-conclusion}

By replacing the diagonal matrix based on the unneeded parameter $c_2$
with a new unitary matrix that performs both sub-block-based DFT and
symbol mapping, an A$^2$FDM scheme was proposed to mitigate the PAPR
problem conflicted by AFDM. Specifically, two symbol mapping schemes
for A$^2$FDM, yielding IA$^2$FDM and LA$^2$FDM, respectively, were
proposed and investigated. Accordingly, the input-output relationships
of IA$^2$FDM and LA$^2$FDM are derived, based on which the effect of
parameter $\mu$, denoting the number of DFT sub-blocks, is
analyzed. The effect of parameter $c_1$ on the performance of AFDM and
IA$^2$FDM is analyzed to explain that the formula provided in
literature for setting $c_1$ is only a necessary condition, but not
sufficient. Furthermore, a comprehensive set of results were provided
to demonstrate the impacts of involved parameters and to compare the
performance of IA$^2$FDM, LA$^2$FDM and AFDM systems. Our analyses
show that both IA$^2$FDM and LA$^2$FDM allow to circumvent the PAPR
problem of AFDM. Owing to that the PAPR of both IA$^2$FDM and
LA$^2$FDM is on the order of $\mu$, instead of $N$ of the total number
of subcarriers. Hence, the PAPR of A$^2$FDM signals can be configured
according to practical application conditions, such as uplink,
downlink, dynamic processing capability of transmitters,
etc. Furthermore, while softening the PAPR problem of AFDM, the
studies and results show that A$^2$FDM also provides the design
options, which are capable of guaranteeing the minimum diversity gain
on the order of $\mu$, and hence avoiding the loss of all the
diversity gain in AFDM, in the cases when parameter $c_1$ happens to
be a `bad' value for some channel conditions.

Our future research issues on the topic will include investigating the other candidate symbol mapping schemes and their consequences, as well as studying A$^2$FDM with integrated sensing and communications
(ISAC).

\begin{appendices}
  \section{Derivation of \eqref{eq x'n summary}}\label{Appendix-A2FDM-1}
 Upon substituting \eqref{eq elements form of z} into $x'_{(b\mu+a)}$,  which is shown above \eqref{eq x'n summary}, we obtain
\begin{equation}
    \begin{aligned}
        x'_{(b\mu+a)} &= \frac{1}{\sqrt{N_\mu}}\sum^{N_\mu-1}_{p=0} z_{kN_\mu+p}  e^{j2\pi\frac{p(b\mu+a)}{N}}\\
        &=\frac{1}{\sqrt{N_\mu}}\sum^{N_\mu-1}_{p=0} \frac{1}{\sqrt{N_\mu}}\sum^{N_\mu-1}_{l=0}s_{k N_\mu+l}e^{-j2\pi\frac{pl}{N_\mu}}  e^{j2\pi\frac{p(b\mu+a)}{N}}\\
        &=\frac{1}{N_\mu}\sum^{N_\mu-1}_{l=0}s_{k N_\mu+l}\sum^{N_\mu-1}_{p=0}e^{j2\pi\frac{((b\mu+a)-l \mu)p}{N}}
    \end{aligned}
    \label{eq x' LDA2FT}
\end{equation}
From \eqref{eq x' LDA2FT} we have that, if $a=0$, $n = b\mu$, and then $x'_n = x'_{(b\mu+a)}$ can be simplified to
    \begin{equation}
    \begin{aligned}
        x'_n = x'_{b\mu} &=\frac{1}{N_\mu}\sum^{N_\mu-1}_{l=0}s_{k N_\mu+l}\sum^{N_\mu-1}_{p=0}e^{j2\pi\frac{(b-l)\mu p}{N}}\\
        &=\frac{1}{N_\mu}\sum^{N_\mu-1}_{l=0}s_{k N_\mu+l}\sum^{N_\mu-1}_{p=0}e^{j2\pi\frac{(b-l) p}{N_\mu}}
    \end{aligned}
    \label{eq x'n, a=0 LDA2FT-1}
\end{equation}
Express $S_{bl} = \sum^{N_\mu-1}_{p=0}e^{j 2\pi\frac{(b-l) p}{N_\mu}}$ and let $\psi = e^{j2\pi\frac{(b-l)}{N_\mu}}$. Explicitly, if $b=l$, then $\psi = 1$ and $S_{bl}=N_\mu$. By contrast, if $b\neq l$ and $\psi\neq 1$, we have $S_{bl} = \sum^{N_\mu-1}_{p=0}\psi^p=
\frac{1-\psi^{N_\mu}}{1-\psi}=0$.  Expressing the above-mentioned in a compact form, if $a=0$, we have
\begin{equation}
    S_{bl} = \sum^{N_\mu-1}_{p=0}e^{j 2\pi\frac{(b-l) p}{N_\mu}} = 
    \begin{cases}
N_\mu, & b=l\text{ }(\text{or }\psi=1), \\
0, & b\neq l\text{ }(\text{or }\psi\neq 1).
\end{cases}
    \label{summation formula for a geometric series}
\end{equation}
Accordingly, $x'_{b\mu}$ can be written as
\begin{equation}
    \begin{aligned}
        x'_{b\mu} &=\frac{1}{N_\mu}\sum^{N_\mu-1}_{l=0}s_{k N_\mu+l}\times S_{bl}\\
        & = \frac{1}{N_\mu}\left(s_{k N_\mu+b} \times N_\mu+\sum_{l=0,l\neq b}^{N_\mu-1}s_{kN_\mu+b} \times 0\right)\\
        &=s_{k N_\mu+b}.
    \end{aligned}
    \label{eq x', n=lmu LDA2FT-2}
\end{equation}

In the case of  $a\neq 0$,  $x'_n$ can be simplified as
\begin{equation}
    \begin{aligned}
    x'_n&=\frac{1}{N_\mu}\sum^{N_\mu-1}_{l=0}s_{k N_\mu+l}\left(\frac{1-e^{j\frac{2\pi(n-l\mu)N_\mu}{N}}}{1-e^{j\frac{2\pi(n-l\mu)}{N}}}\right)\\
    &= \frac{1}{N_\mu}\sum^{N_\mu-1}_{l=0}s_{k N_\mu+l}\left(\frac{1-e^{j2\pi(\frac{n}{\mu}-l)}}{1-e^{j2\pi(\frac{n}{N}-\frac{l}{N_\mu})}}\right)\\
&= \frac{1}{N_\mu}\sum^{N_\mu-1}_{l=0}s_{k N_\mu+l}\left(\frac{1-e^{j\frac{2\pi n}{\mu}}}{1-e^{j\frac{2\pi n}{N}}e^{-j\frac{2\pi l}{N_\mu}}}\right).\\
    \end{aligned}
    \label{eq x', n no equal to lmu LDA2FT}
\end{equation}

Consequently, when both $a=0$ and $a\neq 0$ are considered, $x_n'$ can be represented as \eqref{eq x'n summary}.
\end{appendices}

%\bibliography{A2FDM-references}

\end{document}